\documentclass[11pt]{article}

\usepackage{fullpage}
\usepackage[ruled, linesnumbered, vlined, commentsnumbered]{algorithm2e}
\usepackage{graphicx,subfig} 
\usepackage{amsfonts,amssymb,amsmath,amsthm,amsopn}	
\usepackage{hhline,booktabs,colortbl,multirow,tabularx,threeparttable,makecell} 
\usepackage[listings,skins,breakable]{tcolorbox}

\usepackage{enumerate}
\usepackage{authblk}
\usepackage{footnote}
\usepackage{hyperref}
\usepackage{prettyref}
\usepackage{cite}
\usepackage{setspace}
\usepackage{color}
\usepackage{xcolor}  
\usepackage{scrpage2}
\usepackage{geometry}

\usepackage{tikz}
\usepackage{pgfplots}
\usetikzlibrary{positioning,shapes,shadows,arrows,calc}
\tikzstyle{component}=[rectangle, draw=black, rounded corners, fill=blue!40, drop shadow, text centered, anchor=north, text=white, minimum height=1cm]
\tikzstyle{arrow}=[->, thick]

\pgfplotsset{compat=1.12}
\usetikzlibrary{intersections}
\usetikzlibrary{pgfplots.statistics}
\usepgfplotslibrary{fillbetween}

\def\hlinew#1{%
  \noalign{\ifnum0=`}\fi\hrule \@height #1 \futurelet
   \reserved@a\@xhline}
\makeatother

\definecolor{myblue}{RGB}{34,31,217}
\definecolor{mycyan}{gray}{.7}
\definecolor{Gray}{gray}{0.9}

\DeclareMathOperator*{\argmax}{argmax}

\newcommand{\pref}{\prettyref}

\newrefformat{fig}{Fig.~\ref{#1}}
\newrefformat{tab}{Table~\ref{#1}}
\newrefformat{sec}{Section~\ref{#1}}
\newrefformat{app}{Appendix~\ref{#1}}
\newrefformat{alg}{Algorithm~\ref{#1}}
\newrefformat{property}{Property~\ref{#1}}
\newrefformat{theorem}{Theorem~\ref{#1}}
\newrefformat{corollary}{Corollary~\ref{#1}}
\newrefformat{proposition}{Proposition~\ref{#1}}
\newrefformat{def}{Definition~\ref{#1}}
\newrefformat{eq}{equation~(\ref{#1})}

\title{\vspace{-1ex}\LARGE\textbf{BiLO-CPDP: Bi-Level Programming for Automated Model Discovery in Cross-Project Defect Prediction}\footnote{This manuscript is accepted for publication in ASE 2020. The copyright of this paper has been permanently transferred to ACM.}}

\author[1]{\normalsize Ke Li}
\author[2]{\normalsize Zilin Xiang}
\author[3]{\normalsize Tao Chen}
\author[4]{\normalsize Kay Chen Tan}
\affil[1]{\normalsize Department of Computer Science, University of Exeter, EX4 4QF, Exeter, UK}
\affil[2]{\normalsize College of Computer Science and Engineering, University of Electronic Science and Technology of China, 611731, Chengdu, China}
\affil[3]{\normalsize Department of Computer Science, Loughborough University, Loughborough, LE11 3TU, UK}
\affil[4]{\normalsize Department of Computer Science, City University of Hong Kong, Tat Chee Avenue, Hong Kong}
\affil[$\ast$]{\normalsize Email: \texttt{k.li@exeter.ac.uk, zilin.xiang@gmail.com, t.t.chen@lboro.ac.uk}}

\date{}

\begin{document}
\maketitle

\vspace{-3ex}
{\normalsize\textbf{Abstract: }Cross-Project Defect Prediction (CPDP), which borrows data from similar projects by combining a transfer learner with a classifier, have emerged as a promising way to predict software defects when the available data about the target project is insufficient. However, developing such a model is challenge because it is difficult to determine the right combination of transfer learner and classifier along with their optimal hyper-parameter settings. In this paper, we propose a tool, dubbed \texttt{BiLO-CPDP}, which is the first of its kind to formulate the automated CPDP model discovery from the perspective of bi-level programming. In particular, the bi-level programming proceeds the optimization with two nested levels in a hierarchical manner. Specifically, the upper-level optimization routine is designed to search for the right combination of transfer learner and classifier while the nested lower-level optimization routine aims to optimize the corresponding hyper-parameter settings. To evaluate \texttt{BiLO-CPDP}, we conduct experiments on 20 projects to compare it with a total of 21 existing CPDP techniques, along with its single-level optimization variant and \texttt{Auto-Sklearn}, a state-of-the-art automated machine learning tool. Empirical results show that \texttt{BiLO-CPDP} champions better prediction performance than all other 21 existing CPDP techniques on 70\% of the projects, while being overwhelmingly superior to \texttt{Auto-Sklearn} and its single-level optimization variant on all cases. Furthermore, the unique bi-level formalization in \texttt{BiLO-CPDP} also permits to allocate more budget to the upper-level, which significantly boosts the performance.}



\section{Introduction}
\label{sec:introduction}

Software defects are errors in code and its logic that cause a software product to malfunction or to produce incorrect/unexpected results. Given that software systems become increasingly ubiquitous in our modern society, software defects are highly likely to result in disastrous consequences to businesses and daily lives. For example, the latest \textit{Annual Software Fail Watch} report from Tricentis\footnote{https://www.tricentis.com/resources/software-fail-watch-5th-edition/} shows that, globally, software defects/failures affected over 3.7 billion people and caused \$1.7 trillion in lost revenue. 

One of the key reasons behind the prevalent defects in modern software systems is their increasingly soaring size and complexity. Due to the limited resource for software quality assurance and the intrinsic dependency among a large number of software modules, it is expensive, if not impossible, to rely on human efforts (e.g., code review) to thoroughly inspect software defects. Instead, it is more pragmatic to predict the defect-prone software modules to which software engineers are suggested to focus their limited software quality assurance resource. To this end, machine learning algorithms have been widely used to automate the process of defect prediction.

As discussed in~\cite{ZimmermannNGGM09}, one of the keys to the success of defect prediction models is the amount of data available for model training. In practice, however, it is unfortunately not uncommon that such data is scarce or even unavailable. This can be attributed to the small size of the company and/or the targeted software project is the first of its kind. Cross project defect prediction (CPDP), which aims to predict defects in the a software project by leveraging experience (e.g., training data or hyper-parameters of trained defect prediction models) from other existing ones, has therefore become extremely appealing~\cite{HosseiniTG19}. Unfortunately, partially due to the difference of the data distribution between the source and the target projects, the performance of vanilla CPDP is not as promising as it was supposed to be~\cite{RahmanPD12}. Transfer learning, which is able to transfer knowledge across different domains, has shown to be able to overcome the aforementioned challenges (e.g., data scarcity and data distribution discrepancy) and has gradually become the main driving force for CPDP~\cite{NamPK13}. Generally speaking, the basic idea is to equip a machine learning classifier with a transfer learner that enables its ability to learn from other projects in model building.

There is \textit{No Free Lunch} in defect prediction given that machine learning enabled defect prediction models often come with configurable and adaptable parameters (87\% prevalent classifiers are with at least one parameter~\cite{Tantithamthavorn16,Tantithamthavorn19}). The prediction accuracy on various software projects largely depends on the parameter settings of those defect prediction models~\cite{MendeK09,Mende10}. Furthermore, it becomes more complicated in CPDP because: 1) the configurable parameters is augmented by the transfer learner (85\% widely used CPDP techniques require at least one parameter to setup in the transfer learner) thus lead to an enlarged search space; 2) there exist complex yet unknown interactions among the parameters of the classifier and those of the transfer learner (that is to say parameter optimization over either the classifier or the transfer learner alone may not lead to the overall optimal performance); and 3) the optimal selection of the combination of classifier and transfer learner is as important as parameter optimization but is unfortunately ignored in the current literature (most, if not all, CPDP models are designed with an \textit{ad-hoc} combination of transfer learner and classifier, the performance of which is reported to be far from optimal~\cite{LiXCWT20}). Although there exist some prior works considering the hyper-parameter optimization for CPDP models~\cite{QuCZJ18,Ozturk19}, they only consider the hyper-parameters associated with the classifier. As investigated in a latest empirical study~\cite{LiXCWT20}, this practice is far from truly optimizing the performance of the underlying CPDP model while the settings of hyper-parameters of the transfer learner are more decisive.

Bearing the above considerations in mind, we propose a new tool, dubbed \texttt{BiLO-CPDP}, to automate the model discovery for CPDP tasks. It provides an unified perspective for the combinatorial selection of classifier and transfer learner, as well as their hyper-parameter optimization within the mathematical framework of bi-level programming, where two levels of nested optimization problems are formulated: the upper-level optimization problem is solved subject to the optimality of a lower-level optimization problem. Specifically, the upper-level optimization problem aims to identify the optimal combination of transfer learner and classifier from a given portfolio; while the lower-level optimization problem is dedicated to searching for the optimal parameter setting associated with the corresponding transfer learner and classifier. Note that a combination of transfer learner and classifier is not considered to be feasible for comparison unless the corresponding parameters have been optimized. In \texttt{BiLO-CPDP}, the upper-level optimization is formulated as a combinatorial optimization problem which is solved by the Tabu search~\cite{GloverL98} while the lower-level optimization is modeled as an expensive optimization problem with a limited budget to be solved by Tree-structured Parzen Estimator (TPE)~\cite{BergstraBBK11}, a state-of-the-art Bayesian optimization algorithm.

To evaluate the the effectiveness of \texttt{BiLO-CPDP} for automated model discovery in CPDP, we conduct experiments to compare it with 21 existing CPDP techniques, its single-level variant and \texttt{Auto-Sklearn}~\cite{FeurerKESBH15} \textemdash\ a state-of-the-art automated machine learning (AutoML) tool\textemdash\ over 20 distinct projects. The results fully demonstrate the overwhelming superiority of \texttt{BiLO-CPDP} over the others with statistical significance and a large effect size on all projects.

In summary, the key contributions of this paper are as follows:
\begin{itemize}
    \item To the best of our knowledge, \texttt{BiLO-CPDP} is the first of its kind for automating CPDP from the perspective of bi-level programming. Given that \texttt{BiLO-CPDP} is not only able to automatically search the optimal combination of transfer learner and classifier, but also can set their appropriate hyper-parameter settings, it paves a new avenue for automated model discovery in CPDP.
    \item Through extensive experiments with 21 existing CPDP techniques, we show that \texttt{BiLO-CPDP} is the best on 14 out of 20 projects, and second to only one existing technique for another five. This fully demonstrates the effectiveness and importance brought by automatically choosing the appropriate transfer learner and classifier associated with their optimal hyper-parameter settings for CPDP.
    \item In terms of optimization problem formulation, on all projects, we show that the bi-level programming formulated in \texttt{BiLO-} \texttt{CPDP} is statistically better than hybridizing both combinatorial selection and parameter optimization as a single-level global optimization problem, which is perhaps a more conservative solution as used in, e.g., \texttt{Auto-Sklearn}~\cite{FeurerKESBH15}.
    \item Interestingly, from our experimental results, we disclose that choosing the best combination of classifier and transfer learner (upper level) is more important than fully optimizing their parameters (lower level). Henceforth, given the limited resource for software quality assurance, it is beneficial to allocate more search budget to the upper-level optimization.
\end{itemize}

In the rest of this paper, \pref{sec:bilevel_programming} gives the background about bi-level programming.~\pref{sec:bilo_cpdp} delineates the algorithmic implementation of \texttt{BiLO-CPDP} step by step. The experimental setup is introduced in~\pref{sec:evaluation} and the results are analyzed in~\pref{sec:result}. Thereafter, \pref{sec:related} and~\ref{sec:threats} reviews the related works and discusses the threats to validity, respectively. Finally, \pref{sec:conclusion} concludes this paper and threads some lights on future directions.



\section{Bi-level Programming}
\label{sec:bilevel_programming}

Bi-level programming is a mathematical program within which one optimization problem is nested within another in a hierarchical manner~\cite{SinhaMD18}. It is ubiquitous in many real-world optimization and public/private sector decision-making problems where the realized outcome of any solution or decision taken by the upper-level authority (a.k.a. leader) to optimize their objectives is affected by the response of lower-level entities (a.k.a. follower), who seek to optimize their own outcomes. This is in principle similar to the Stackelberg games~\cite{von2010market} in which a leader first makes its move and a follower maximizes the corresponding gain by taking the leader's move into account. It is interesting to note that the two levels of optimization problems are asymmetric in bi-level programming. That is to say, the upper-level leader has the entire picture of optimization problems at both levels whereas the lower-level follower usually takes the decisions from the leader and then optimizes its own strategies.

The bi-level programming formulated in \texttt{BiLO-CPDP} can be mathematically defined as:
\begin{equation}
\label{eq:bilevel}
	\begin{array}{ll}
	{\underset{\mathbf{x}^u\in\Lambda^d\times\mathbb{R}^n,\mathbf{x}^l\in\mathbb{R}^n}{\operatorname{maximize}}} & {F(\mathbf{x}^u, \mathbf{x}^{l\ast})} \\
	{\text { subject to }} & {\mathbf{x}^{l\ast} \in \operatorname{argmax}\{f_{\mathbf{x}^u}(\mathbf{x}^l)\}}
	\end{array},
\end{equation}
where $\mathbf{x}^u\in\Lambda^d\times\mathbb{R}^n$ and $\mathbf{x}^l\in\mathbb{R}^n$ denote the upper- and lower-level variables\footnote{$\Lambda^d\times\mathbb{R}^n$ means that the problem is a discrete combinatorial problem.} while $F:\Lambda^d\times\mathbb{R}^n\rightarrow\mathbb{R}$ and $f_{\mathbf{x}^u}:\mathbb{R}^n\rightarrow\mathbb{R}$ are the upper-level and lower-level objective functions, respectively (details can be found in Section~\ref{sec:bilo_cpdp}). A bi-level programming that involves nested optimization/decision-making tasks at both levels. For any given combination $\mathbf{x}^u$, there exists a $(\mathbf{x}^u,\mathbf{x}^{l\ast})$ pair where $\mathbf{x}^{l\ast}$ is an optimal (or near-optimal) response to $\mathbf{x}^u$ represents a feasible solution to the upper-level optimization problem given that it also satisfies the constraints therein.


\section{Bi-level Programming for Automated CPDP Model Discovery}
\label{sec:bilo_cpdp}

The CPDP model building process consists of two intertwined parts: 1) transfer learning that augments data from different domains by selecting relevant instances or assigning appropriate weights to different instances; and 2) defect prediction model building based on adapted data. As reported in a latest research~\cite{LiXCWT20}, the performance of a CPDP model largely depend on the combination of transfer learner and classifier along with their hyper-parameter settings. In light of this, the \texttt{BiLO-CPDP} proposed in this work was specifically designed to address such a problem. Through automatically discovering the best combination of transfer learner and classifier as well as their optimal hyper-parameter settings, \texttt{BiLO-CPDP} serves as an automatic tool that provides a \textit{de nova} CPDP model discovery. In this section, we will delineate the architecture of \texttt{BiLO-CPDP} and the algorithmic details of its optimization routines at both levels.

\begin{figure}[t]
    \centering
	\includegraphics[width=\linewidth]{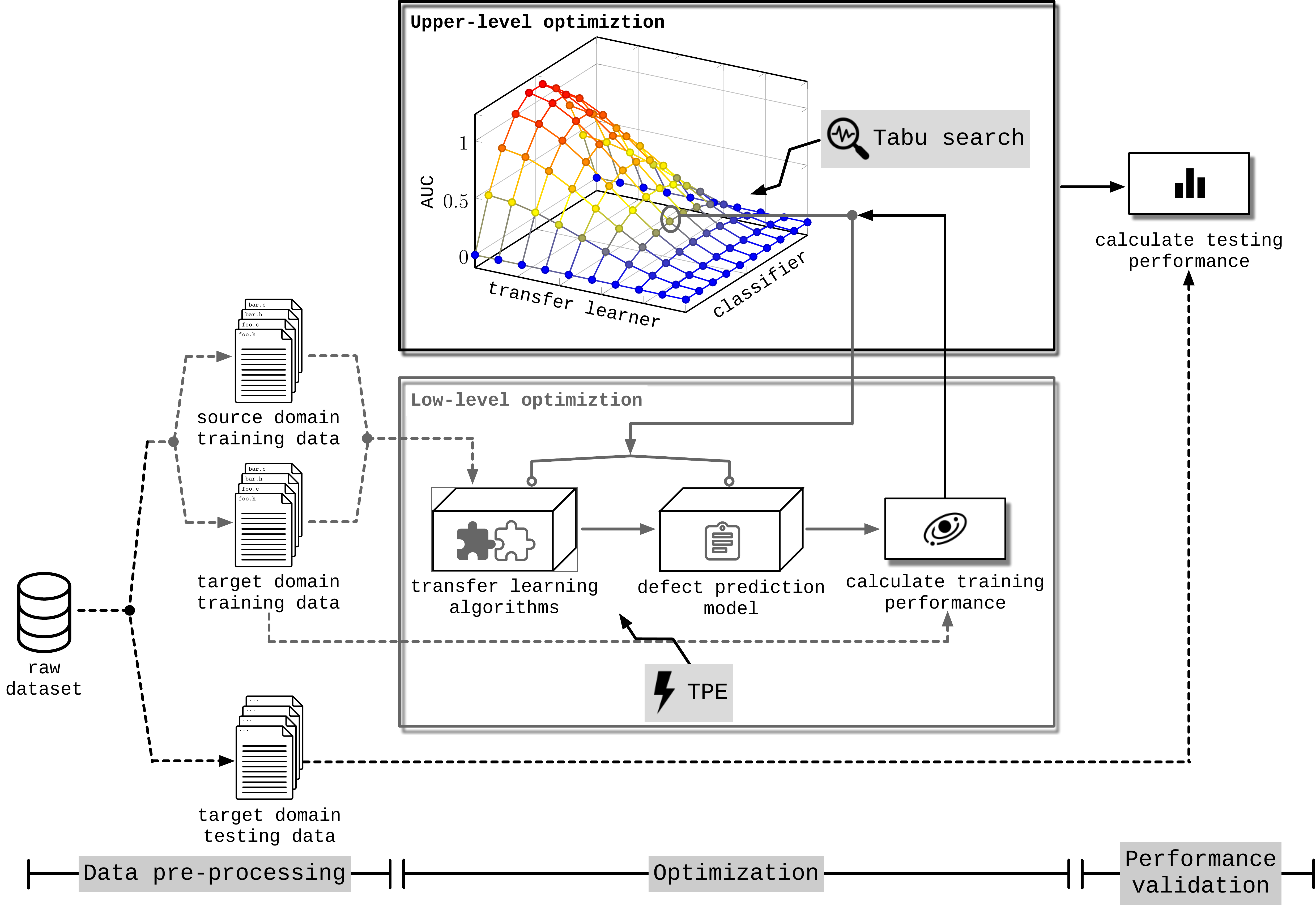}
	\caption{The overall architecture of \texttt{BiLO-CPDP}.}
	\label{fig:BiLO-CPDP}
\end{figure}

\subsection{Overview of \texttt{BiLO-CPDP}}
\label{sec:architecture}

The overall architecture of \texttt{BiLO-CPDP} is illustrated in~\pref{fig:BiLO-CPDP} which consists of three key phases, i.e., \textit{data pre-processing}, \textit{optimization} and \textit{performance validation}.
\begin{enumerate}
    \item\textbf{\underline{Data Pre-processing:}} Given a raw dataset with $N>1$ projects, software engineers are asked to specify which one is the target domain that serves as the target domain data while the remaining $N-1$ projects are then used as the source domain data. In particular, all source domain data are used in the model training while a part of the target domain data is used as the hold-out set for the testing purpose. As the default in \texttt{BiLO-CPDP}, we use 10\% of the target domain data for testing while the remaining 90\% is for training. This is because some transfer learners considered in this work do need data from the target domain in training, e.g., \texttt{MCWs}~\cite{qiu2018multiple}. For other transfer learners that can be trained independently to the target domain, we use all data for testing.
    \item\textbf{\underline{Optimization:}} \texttt{BiLO-CPDP} models CPDP as a bi-level programming that not only identifies the most competitive combination of transfer leaner and classifier for the underlying CPDP task (tackled by the upper-level routine), but also equips the chosen CPDP model with the appropriate hyper-parameter settings (carried out by the lower-level routine). Since the resources for software quality assurance are often limited, the entire optimization process would inevitably be constrained under a computational budget of running time. In this regard, the unique bi-level programming formulated in \texttt{BiLO-CPDP} can in fact provide a fine-grained and flexible allocation of the budget between upper- and lower-level, whose effects will be investigated as part of the experimental evaluation in \pref{sec:two_level_importance}. The CPDP model, which has the best combination with its optimal hyper-parameter settings in terms of the training accuracy, is returned in the end. Note that due to the lack of data samples, using training accuracy in the parameter optimization of transfer learner is not uncommon and has shown promising results for CPDP~\cite{LiXCWT20}.
    \item\textbf{\underline{Performance Validation:}} After the optimization phase, as an optional module in \texttt{BiLO-CPDP}, the generalization of the built CPDP model can be validated and tested by using the hold-out set from the target domain data, which is unknown during training stage. In practice, this will be the new project that one wishes to predict defects for. In \texttt{BiLO-CPDP}, the area under the receiver operating characteristic (ROC) curve, i.e., AUC~\cite{ZhouYLCLZQX18}, is applied as the performance metric to measure the effectiveness of a model. Formally, AUC is defined as: 
    \begin{equation}
		AUC = \frac{\sum_{t^-\in\mathcal{D}^{-}} \sum_{t^{+}\in\mathcal{D}^{+}} \mathbf{1}\left[Pred\left(t^{-}\right)<Pred\left(t^{+}\right)\right]}{\left|\mathcal{D}^{-}\right| \cdot\left|\mathcal{D}^{+}\right|},
	\end{equation}
	where $Pred(t)$ is the probability that sample $t$ is predicted to be a positive sample, and $\mathbf{1}\left[f\left(t^{-}\right)<f\left(t^{+}\right)\right]$ is an indicator function which returns 1 if $ f(t^{-})<f(t^{+})$ otherwise it returns 0. ${\displaystyle {\mathcal {D}}^{-}}$ is the set of negative samples, and ${\displaystyle {\mathcal {D}}^{+}}$ is the set of positive samples. Apart from the fact that AUC has been widely for software defect prediction~\cite{LiXCWT20}, it has two distinctive characteristics: 1) different from other prevalent metrics like precision and recall, AUC does not depend on a particular threshold~\cite{ZhouYLCLZQX18}, which is difficult to tweak in order to carry out an unbiased assessment; and 2) it is not sensitive to imbalanced data which is not uncommon in software defect prediction~\cite{LiJZ18}. The larger the AUC value is, the better prediction accuracy the model achieves. In particular, the AUC value ranges between 0 and 1 where 0 indicates the worst performance, 0.5 corresponds a randomly guessed performance and 1 represents the best performance. Note that AUC is also the metric used in \emph{optimization} phase to evaluate and compare training accuracy.
	

\end{enumerate}

\begin{table*}
\footnotesize
  \caption{Overview of 13 selected transfer learners ([N], [R] and [C] denote integer, real and categorical value, respectively).}
 \label{tab:transfer_learner}
\setlength{\tabcolsep}{1.1mm}
\centering
\begin{tabular}{c|c|c||c|c|c||c|c|c}
\hline

\textbf{Algorithm}&\textbf{Parameter}&\textbf{Range}&\textbf{Algorithm}&\textbf{Parameter}&\textbf{Range}&\textbf{Algorithm}&\textbf{Parameter}&\textbf{Range}\\ 
\hline
\makecell{\texttt{NNfilter}\\\cite{TurhanMBS09}}&\makecell{k [N]\\metric [C]\\\hfill}&\makecell{$[1,100]$\\Euc, Man, Che, \\Min, Mah}&\makecell{\texttt{CDE\_SMOTE}\\\cite{limsettho2018cross}}&\makecell{k [N]\\metric [C]\\\hfill}&\makecell{$[1,100]$\\Euc, Man, Che, \\Min, Mah}&\makecell{\texttt{FSS\_bagging}\\\cite{he2013learning}}&\makecell{topN [N]\\threshold [R]\\ratio [R]}&\makecell{$[1,15]$\\$[0.3,0.7]$\\$[0.1,0.5]$}\\


\hline


\makecell{\texttt{TCA+}~\cite{NamPK13}}&\makecell{kernel [C]\\\hfill \\dime [N]\\\hfill \\lamb [R]}&\makecell{primal, rbf,\\linear, sam\\$[5$, max(N\_s,\\N\_t)]\\$[10E-7,100]$}&\makecell{\texttt{GIS }~\cite{hosseini2018benchmark}}&\makecell{prob [R]\\chrm\_size [R]\\pop\_size [N]\\num\_parts [N]\\num\_gens [N]}&\makecell{$[0.02,0.1]$\\$[0.02,0.1]$\\$[2,30]$\\$[2,6]$\\$[5,20]$}&\makecell{\texttt{CLIFE\_MORPH} \\\cite{peters2013balancing}}&\makecell{n [N]\\alpha [R]\\beta [R]\\ per [R]}&\makecell{$[1,100]$\\$[0.05,0.2]]$\\$[0.2,0.4]$\\$[0.6, 0.9]$}\\

\cline{7-9}

&gama [R]&$[10E-6,100]$&&mcount [N]&$[3,10]$&\makecell{\texttt{HISNN}~\cite{ryu2015hybrid}}&\makecell{minham [N]}&\makecell{$[1$, N\_s]}\\

\hline

\makecell{\texttt{MCWs}~\cite{qiu2018multiple}}&\makecell{k [N]\\sigma [R]\\lambda [R]}&\makecell{$[2$, N\_s]\\$[0.01,10]$\\$[10E-7, 100]$}&\makecell{\texttt{FeSCH}~\cite{ni2017cluster}}&\makecell{nt [N]\\strategy [C]}&\makecell{$[1$, N\_s]\\SFD, LDF, FCR}&\makecell{\texttt{UM} ~\cite{zhang2014towards}}&\makecell{$p$ [R]\\qua\_T [C]}&\makecell{$[0.01,0.1]$\\cli , cohen\\}\\

\hline

\makecell{\texttt{TD}~\cite{herbold2013training}}&\makecell{strategy [C]\\k [N]}&\makecell{NN, EM\\$[1$, N\_s]}&\makecell{\texttt{VCB}~\cite{ryu2016value}}&\makecell{m [N]\\lambda [R]}&\makecell{$[2,30]$\\$[0.5,1.5]$}&\makecell{\texttt{PCAmining}\\\cite{nagappan2006mining}}&\makecell{dime [N]\\\hfill}&\makecell{$[5$, max(N\_s,\\N\_t)\\}\\

\hline
\end{tabular}
 \begin{tablenotes}
    \footnotesize
    \item For full specification of all the parameters, please visit our repository: \url{https://github.com/COLA-Laboratory/ase2020}
    \end{tablenotes}
\end{table*}

\begin{table*}
\footnotesize
	\caption{Overview of 16 selected classifiers ([N], [R] and [C] denote integer, real and categorical value, respectively).}
	\label{tab:classifier}
\setlength{\tabcolsep}{0.8mm}
\centering
\begin{tabular}{c|c|c||c|c|c||c|c|c}
\hline
\textbf{Algorithm}&\textbf{Parameter}&\textbf{Range}&\textbf{Algorithm}&\textbf{Parameter}&\textbf{Range}&\textbf{Algorithm}&\textbf{Parameter}&\textbf{Range}\\ 
\hline

\makecell{\texttt{Extra}\\\texttt{Trees}\\\texttt{Classifier}\\\texttt{ (EXs)}}&\makecell{max\_e [N]\\criterion [C] \\min\_s\_l [N]\\splitter [C]\\min\_a\_p [N]}&\makecell{$[10,100]$\\gini, entropy\\$[1,20]$\\random, best\\$[2$, N\_s/10]}&\makecell{\texttt{Extra}\\\texttt{Tree}\\\texttt{Classifier}\\\texttt{(EXtree)}}&\makecell{max\_e [N]\\criterion [C] \\min\_s\_l [N]\\splitter [C]\\min\_a\_p [N]}&\makecell{$[10,100]$\\gini, entropy\\$[1,20]$\\random, best\\$[2$, N\_s/10]}&\makecell{\texttt{Decision}\\\texttt{Tree (DT)}}&\makecell{max\_e [N]\\criterion [C]\\min\_s\_l [N]\\splitter [C]\\min\_a\_p [N]}&\makecell{$[10,100]$\\gini, entropy\\$[1,20]$\\auto, sqrt, log2\\$[2$, N\_s/10]}\\
\hline

\makecell{\texttt{Random}\\\texttt{Forest (RF)}}&\makecell{m\_stim [N]\\criterion [C]\\splitter [C]\\min\_s\_l [N]\\min\_a\_p [N]}&\makecell{$[10,100]$\\gini, entropy\\auto, sqrt, log2\\$[1,20]$\\$[2$, N\_s/10]}&\makecell{\texttt{Support}\\\texttt{Vector}\\\texttt{Machine}\\\texttt{(SVM)}}&\makecell{C [R]\\kernel [C]\\degree [N]\\coef0 [R]\\gamma [R]}&\makecell{$[0.001,10]$\\rbf, lin, poly, sig\\$[1,5]$\\$[0,10]$\\$[0.01, 100]$}&\makecell{\texttt{Multilayer}\\\texttt{Perceptron}\\\texttt{(MLP)}}&\makecell{active [C]\\\hfill \\hid\_l\_s [N]\\ solver [C]\\ iter [N]}&\makecell{iden, log,\\tanh, relu\\$[50, 200]$\\lbfgs, sgd, adam\\$[100,250]$}\\


\hline

\makecell{\texttt{Passive}\\\texttt{Aggressive}\\\texttt{Classifier}\\\texttt{(PAC)}}&\makecell{C [R]\\fit\_int [C]\\tol [R]\\loss [C]}&\makecell{$[0.001,100]$\\true, false\\$[10E-6,0.1]$\\hinge, s\_hinge}&\makecell{\texttt{Perceptron}}&\makecell{penalty [C]\\ alpha [R]\\fit\_int [C]\\tol [R]}&\makecell{L1, L2\\$[10E-6,0.1]$\\true, false\\$[10E-6,0.1]$}&\makecell{\texttt{Naive}\\\texttt{Bayes (NB)}}&\makecell{NBType [C]\\\hfill\\alpha [R]\\ norm [C]}&\makecell{gauss,\\ multi, comp\\$[0, 10]$\\ture, false}\\

\hline

\makecell{\texttt{Ridge}}&\makecell{alpha [R]\\fit\_int [C]\\tol [R]}&\makecell{$[10E-5,1000]$\\ture, false\\$[10E-6,0.1]$}&\makecell{\texttt{Bagging}}&\makecell{n\_est [N]\\max\_s [R]\\max\_f [R]}&\makecell{$[10,200]$\\$[0.7,1.0]$\\$[0.7,1.0]$}&\makecell{\texttt{Logistic}\\\texttt{Regression}\\\texttt{(LR)}}&\makecell{penalty [C]\\fit\_int [C]\\tol [R]}&\makecell{L1, L2\\ture, false\\$[10E-6,0.1]$}\\

\hline


\makecell{\texttt{KNearest N-}\\\texttt{eighbor(KNN)}}&\makecell{n\_neigh [N]\\p [N]}&\makecell{$[1, 50]$\\$[1, 5]$}&\makecell{\texttt{Radius}\\\texttt{Neighbors}}&\makecell{radius [R]\\weight [C]}&\makecell{$[0,10000]$\\uni, dist}&\makecell{\texttt{Nearest}\\\texttt{Centroid}}&\makecell{metric [C]\\\hfill}&\makecell{Euc, Man, Che,\\Min, Mah}\\

\cline{1-3}

\makecell{\texttt{adaBoost}}&\makecell{n\_est [N]\\rate [R]}&\makecell{$[10,1000]$\\$[0.01,10]$}&\makecell{\texttt{Classifier}\\\texttt{(RNC)}}&&&\makecell{\texttt{Classifier}\\\texttt{(NCC)}}&\makecell{shrink\_t [R]\\\hfill}&\makecell{$[0,10]$\\\hfill}\\

\hline
\end{tabular}
 \begin{tablenotes}
    \footnotesize
    \item For full specification of all the parameters, please visit our repository: \url{https://github.com/COLA-Laboratory/ase2020}
    \end{tablenotes}
\end{table*}



\subsection{Upper-Level Optimization}
\label{sec:upper_level}

Tables~\ref{tab:transfer_learner} and \ref{tab:classifier} respectively list the transfer learners and the classifiers considered in our work, which form the portfolios. Note that all transfer learners considered in \texttt{BiLO-CPDP} have been used in either the defect prediction or CPDP literature while the classifiers come from \texttt{scikit-learn}\footnote{https://scikit-learn.org/stable/}, the state-of-the-art machine learning Python library. In addition, the corresponding hyper-parameters associated with those transfer learners and classifiers along with their value ranges are also provided in the corresponding tables. Any combination of a transfer learner and a classifier comes up with a CPDP model. The ultimate goal of the upper-level optimization is to search for the best combination out of all possible alternatives (208 in this work) for the underlying CPDP task. In particular, for each candidate combination of transfer learner and classifier, their corresponding hyper-parameter settings are optimized via a lower-level optimization routine which will be explained in~\pref{sec:lower_level}.

At the upper-level in \texttt{BiLO-CPDP}, the search of the best combination of transfer learner and classifier is solved as a combinatorial optimization problem as specified below.
\begin{itemize}
    \item\textbf{\underline{Search space}:} For the upper level, the search space consists of all the valid combinations of transfer learners and classifier picked up from the given portfolios, i.e., those listed in Tables~\ref{tab:transfer_learner} and~\ref{tab:classifier}. In practice, such portfolios can be amended and specified by the software engineers based on their preferences/requirements.
    
    \item\textbf{\underline{Objective function}:} Recall from the \pref{eq:bilevel}, the objective function for the upper level $F(\mathbf{x}^u, \mathbf{x}^{l\ast})$ takes a combination from the portfolio ($\mathbf{x}^u$) and the optimized hyper-parameter of such combination ($\mathbf{x}^{l\ast}$) as inputs. It then outputs the corresponding training AUC obtained by training the CPDP model for comparison. Note that $\mathbf{x}^{l\ast}$ is initially unknown for a given $\mathbf{x}^u$ at the upper-level before running a lower-level optimization routine. Therefore, the objective function at upper-level optimization is constrained and determined by the lower-level optimization.

    \item\textbf{\underline{Optimization algorithm}:} For the upper-level optimization in \texttt{BiLO-CPDP}, we use Tabu search~\cite{GloverL98} to serve as the optimizer, which is also the entry point of the optimization phase. In particular, we use Tabu search in this work because:
    
    \begin{itemize}
        \item Our problem is expensive and thus it is unrealistic for an exact search to reach the optimal solution. Metaheuristic such as Tabu search, which does not guarantee optimum but can often produce near-optimal result, is more practical and acceptable.
        \item Unlike other metaheuristics, Tabu search employs local search to speed up its convergence~\cite{GloverL98}.
        \item Tabu search permits a better chance to escape from local optima than other local search methods~\cite{GloverL98}.
    \end{itemize}

    As shown in in~\pref{alg:bilo} and~\pref{alg:sample_candidate}, Tabu search carries out a neighborhood search where the neighborhood of the current solution is restricted by the search history of previously visited solutions and is stored in the form of a \textit{tabu list} (lines 5 and 6 in~\pref{alg:bilo} and lines 5 to 9 in~\pref{alg:sample_candidate}). If all neighbors are \textit{tabu}, it is acceptable to take a move that worsen the value of the objective function (lines 3 and 4 in~\pref{alg:sample_candidate}). This is what enables Tabu search to escape from local optima, which is highly likely to cause issues with a traditional gradient decent method. According to a provided selection criteria, Tabu search only keep a record of some previously visited states.
\end{itemize}

    \begin{algorithm}[t]
    \caption{\textsc{runTabuSearch}: Upper-level optimization that tunes the combination of transfer learner and classifier.}
    \label{alg:bilo}
    \KwIn{Portfolio of transfer learners $\mathcal{T}$ and classifiers $\mathcal{C}$}
    \KwOut{Optimal CPDP model $\mathbf{x}_\texttt{opt}^u$ and its optimal parameter settings $\mathbf{x}_\texttt{opt}^{l\ast}$}
    Randomly initialize a valid combination of transfer learner and classifier $\mathbf{x}^u\leftarrow(t,c)$\tcc*[r]{$t\in\mathcal{T}$ and $c\in\mathcal{C}$ are a candidate transfer learner and classifier}
    $\ell_t\leftarrow\emptyset$\tcc*[r]{$\ell_t$ is the tabu list}
    \While{The overall time budget is not exhausted}
    {  
        $[(\mathbf{x}^u,\mathbf{x}^{l\ast}),F(\mathbf{x}^u,\mathbf{x}^{l\ast})]\leftarrow\textsc{searchCandidate}(\mathbf{x}^u,\ell_t)$;\\
        \If{$\mathbf{x}^u\notin\ell_t$}{
            $\ell_t\leftarrow\ell_t\bigcup\{\mathbf{x}^u\};$\\
        }
    }
    \Return $(\mathbf{x}_\texttt{opt}^u,\mathbf{x}_\texttt{opt}^{l\ast})\leftarrow\argmax_{\mathbf{x}^u\in\ell_t}\{F(\mathbf{x}^u,\mathbf{x}^{l\ast})\}$;
\end{algorithm}

    \begin{algorithm}[t!]
    \caption{$\textsc{searchCandidate}(\mathbf{x}^u,\ell_t)$: Search the next combination candidate from the neighborhood of $\mathbf{x}^u$}
    \label{alg:sample_candidate}
    \KwIn{Candidate CPDP model $\mathbf{x}^u$ and newest \textit{tabu list} $\ell_t$}
    \KwOut{The best CPDP model within $\mathbf{x}^u$'s neighbor and its optimal parameter settings $\mathbf{x}_\texttt{opt}$, the performance of this CPDP model $f_\texttt{opt}$, i.e., the value of $F(\mathbf{x}^u,\mathbf{x}^{l\ast})$}
    $\delta\leftarrow$ Get the neighbors of $\mathbf{x}^u$;\\
    $\Theta_{\mathbf{x}^u}\leftarrow$ Get the configuration space of the transfer learner and the classifier specified by $\mathbf{x}^u$;\\
    $[\mathbf{x}^{l\prime},f_{\mathbf{x}^u}(\mathbf{x}^{l\prime})]\leftarrow\textsc{runTPE}(\mathbf{x}^u,\Theta_{\mathbf{x}^u})$;\\
    $\mathbf{x}_\texttt{opt}\leftarrow(\mathbf{x}^u,\mathbf{x}^{l\prime})$, $f_\texttt{opt}\leftarrow f_{\mathbf{x}^u}(\mathbf{x}^{l\prime})$;\\
    \ForEach{$\mathbf{x}^c\in\delta\land\mathbf{x}^c\notin\ell_t$}  
    {
        $\Theta_{\mathbf{x}^c}\leftarrow$ Get the configuration space of the transfer learner and the classifier specified by $\mathbf{x}^c$;\\
        $[\mathbf{x}^{l\prime},f_{\mathbf{x}^c}(\mathbf{x}^{l\prime})]\leftarrow\textsc{runTPE}(\mathbf{x}^c,\Theta_{\mathbf{x}^c})$;\\
        \If{$f_{\mathbf{x}^c}(\mathbf{x}^{l\prime})>f_\texttt{opt}$}{
            $\mathbf{x}_\texttt{opt}\leftarrow(\mathbf{x}^c,\mathbf{x}^{l\prime})$, $f_\texttt{opt}\leftarrow f_{\mathbf{x}^c}(\mathbf{x}^{l\prime})$;
        }
    }  
    \Return $\mathbf{x}_\texttt{opt},f_\texttt{opt}$;  
\end{algorithm}

\subsection{Lower-Level Optimization}
\label{sec:lower_level}

As introduced in~\pref{sec:architecture}, the main purpose of the lower-level optimization is to optimize the hyper-parameters associated with the chosen combination of transfer learner and classifier. Specifically, this level in \texttt{BiLO-CPDP} is modeled and tackled as below.

\begin{itemize}
    \item\textbf{\underline{Search space}:} At this level, the search space is the configuration space of the corresponding parameters for the transfer learner and classifier picked up from the upper-level routine. Indeed, as can be seen from Tables~\ref{tab:transfer_learner} and~\ref{tab:classifier}, such a configuration space might be different depending on the chosen combination of transfer learner and classifier.

    \item\textbf{\underline{Objective function}:} Recall from the~\pref{eq:bilevel}, when a combination of transfer learner and classifier is picked up from the upper-level routine, the objective function for the lower-level $f(\mathbf{x}^l)$ takes the configuration of the corresponding hyper-parameters as the inputs ($\mathbf{x}^l$) and outputs the training AUC for the CPDP model. The AUC collected from the result of the low-level routine is finally used as the objective value at the upper-level routine to steer the optimization.
 
    \item\textbf{\underline{Optimization algorithm}:} It is not uncommon that the training and evaluation of a CPDP model is computationally demanding and time consuming. To this end, in \texttt{BiLO-CPDP}, we apply the Tree-structured Parzen Estimator (TPE)~\cite{BergstraBBK11} \textemdash\ a state-of-the-art Bayesian optimization algorithm for hyper-parameter optimization of machine learning algorithms \textemdash\ as the optimizer for the lower-level optimization, due primarily to the following reasons: 
    
    \begin{itemize}
        \item TPE copes with a wide range of variables well, including integer, real, and categorical ones, which fits precisely with our need~\cite{BergstraBBK11}.
        \item Recent work on CPDP~\cite{LiXCWT20} and from the machine learning community~\cite{FeurerH19} have reported the outstanding performance of TPE for expensive configuration problems.
    \end{itemize}

    As the pseudo-code shown in~\pref{alg:lower_opt}, the \texttt{TPE} algorithm first uses a space-filling technique to sample a set of hyper-parameters' values from the given configuration space $\Theta_c$ of transfer learner and classifier, which would then be trained for collecting the training AUC performance (line 1). All these constitute the initial dataset $\mathcal{D}$. During the main while-loop, a relatively cheap surrogate model of the expensive physical model training and the AUC evaluation is built based on all sampled data in $\mathcal{D}$ (line 3). Thereafter, a promising hyper-parameter configuration trial $
    \mathbf{x}^{l_c}$ is identified by optimizing the acquisition function (i.e., expected improvement) following a classic Bayesian optimization rigour. The AUC of $\mathbf{x}^{l_c}$ is thereafter evaluated and used to augment $\mathcal{D}$ (lines 4 to 6). At the end, the best hyper-parameter setting $\mathbf{x}^{l\ast}$ in $\mathcal{D}$ along with its AUC performance $f_{\mathbf{x}^u}(\mathbf{x}^{l\ast})$ are returned to the upper-level optimization routine (line 7).
\end{itemize}

   \begin{algorithm}[t]
    \caption{$\textsc{runTPE}(\mathbf{x}^u,\Theta_{\mathbf{x}^u})$: Lower-level optimization for identifying the optimal hyper-parameters.}
    \label{alg:lower_opt}
    \KwIn{Combination of transfer learner and classifier $\mathbf{x}^u$, configuration space $\Theta_{\mathit{c}}$}
    \KwOut{Optimized hyper-parameters $\mathbf{x}^{l\ast}$ and its objective function $f_{\mathbf{x}^u}(\mathbf{x}^{l\ast})$}
    $\mathcal{D}\leftarrow$ Use space-filling to sample a set of hyper-parameters from $\Theta_c$ and evaluate their objective functions;\\
    \While{The lower-level time budget is not exhausted}  
    {
        Use Tree Parzen to build a surrogate model based on $\mathcal{D}$;\\
        $\mathbf{x}^{l_c}\leftarrow$ Best configuration based on the AUC predicted by the acquisition function over the surrogate model;\\
        $f_{\mathbf{x}^u}(\mathbf{x}^{l_c})\leftarrow$ Evaluate the objective function of $\mathbf{x}^{l_c}$ by physically training the CPDP model;\\
        $\mathcal{D} = \mathcal{D}\bigcup\{(\mathbf{x}^{l_c},f_{\mathbf{x}^u}(\mathbf{x}^{l_c}))\}$;\\
    }
    \Return $(\mathbf{x}^{l\ast},f_{\mathbf{x}^u}(\mathbf{x}^{l\ast}))\leftarrow\argmax_{(\mathbf{x}^{l_c},f_{\mathbf{x}^u}(\mathbf{x}^{l_c}))\in\mathcal{D}}\{f_{\mathbf{x}^u}(\mathbf{x}^{l_c})\}$;
    
\end{algorithm}


\section{Experimental Setup}
\label{sec:evaluation}

This section introduces our experiment setups\footnote{All source code and data of this work can be publicly accessed via our repository: \url{https://github.com/COLA-Laboratory/ase2020}}.

\subsection{Dataset}
\label{sec:dataset}

In our experiments, the dataset of software projects is collected according to the following three \textit{inclusion} criteria:
\begin{enumerate}
	\item To promote the reproducibility and practicality of our experiments, we only consider projects hosted in public repositories and are related to non-academic software.
	\item To mitigate potential conclusion bias, projects are required to cover different corpora and domains. 
	\item To ensure the credibility of experiments, we focus on projects that have already been used in the CPDP literature.
\end{enumerate}
Note that a project is temporarily selected if it meets all above three criteria. To further refine our dataset composition, we apply the following two \textit{exclusion} criteria to rule out inappropriate projects.
\begin{enumerate}
	\item It is not uncommon that the projects are evolved with more than one version during their lifetime. Since different versions of the same project are highly likely to share many similarities, they may simplify the transfer learning. In this case, only the latest version of the project is kept.

	\item To promote the robustness of experiments, projects with repeated or missing data are ruled out from our consideration.
\end{enumerate}

Based on the above inclusion criteria, we select five publicly available datasets, i.e., \texttt{JURECZKO}, \texttt{NASA}, \texttt{SOFTLAB}, \texttt{AEEEM}, \texttt{ReLink}. Note that all these datasets have been reviewed and discussed in many recent survey in the CPDP literature~\cite{Herbold17,HerboldTG18,ZhouYLCLZQX18,HosseiniTG19}. Thereafter, \texttt{SOFTLAB} is further ruled out from our consideration according to the above exclusion criteria. In addition, \texttt{NASA} is also not considered in our experiments since its data quality is relatively poor as reported in~\cite{ShepperdSSM13}. At the end, the dataset considered in our experiments consist of 20 open source projects with 10,952 instances. Its characteristics are summarized as follows:
\begin{itemize}
	\item\texttt{AEEEM}~\cite{DAmbrosLR10}: This dataset contains 5 open source projects with 5,371 instances. In particular, each instance has 61 metrics with two different types, including static and process metrics like the entropy of code changes and source code chorn.
	\item\texttt{ReLink}~\cite{WuZKC11}: This dataset consists of 3 open source projects with 649 instances. In particular, each instance comes with 26 static metrics. Note that the defect labels are further manually verified after being generated from source code management system commit comments.
	\item\texttt{JURECZKO}~\cite{JureczkoM10}: This dataset originally consists of 92 released software collected from a mix of open sourced, proprietary and academic projects. With respect to the first inclusion criterion, those proprietary and academic projects are not considered. Moreover, since the projects in \texttt{JURECZKO} have been updated more than once, according to the first exclusion criterion, only the latest version of a project is considered in our experiments. Ultimately, we choose 12 open source projects with 4,932 instances from \texttt{JURECZKO}.
\end{itemize}

\begin{table*}[t!]   
    \small
    \centering
    \caption{Scott-Knott test on \texttt{BiLO-CPDP} and existing CPDP techniques over 30 runs. (the larger rank, the better; gray=the best)}
	\label{tab:rq1}
    \begin{tabular}{c|cccccccccccccccccccc}
    \hline
          \textbf{CPDP Technique} & \rotatebox{90}{\textbf{Apache}} & \rotatebox{90}{\textbf{EQ}} & \rotatebox{90}{\textbf{JDT}} & \rotatebox{90}{\textbf{LC}} & \rotatebox{90}{\textbf{ML}} & \rotatebox{90}{\textbf{PDE}} & \rotatebox{90}{\textbf{Safe}} & \rotatebox{90}{\textbf{Tomcat}} & \rotatebox{90}{\textbf{Zxing}} & \rotatebox{90}{\textbf{ant}} & \rotatebox{90}{\textbf{camel}} & \rotatebox{90}{\textbf{ivy}} & \rotatebox{90}{\textbf{jEdit}} & \rotatebox{90}{\textbf{log4j}} & \rotatebox{90}{\textbf{lucene}} & \rotatebox{90}{\textbf{poi}} & \rotatebox{90}{\textbf{synapse}} & \rotatebox{90}{\textbf{velocity}} & \rotatebox{90}{\textbf{xalan}} & \rotatebox{90}{\textbf{xerces}} \\
          
    \hline
    \texttt{NNfilter-NB} & \cellcolor[rgb]{ .749,  .749,  .749}\textbf{6} & 5     & 4     & \cellcolor[rgb]{ .749,  .749,  .749}\textbf{8} & \cellcolor[rgb]{ .749,  .749,  .749}\textbf{5} & 3     & \cellcolor[rgb]{ .749,  .749,  .749}\textbf{8} & \cellcolor[rgb]{ .749,  .749,  .749}\textbf{14} & 4     & 6     & \cellcolor[rgb]{ .749,  .749,  .749}\textbf{7} & 9     & 3     & 8     & \cellcolor[rgb]{ .749,  .749,  .749}\textbf{8} & 6     & 2     & 3     & 8     & 7 \\
    \hline
    \texttt{UM-NB} & 2     & 7     & 6     & 5     & 4     & 3     & 1     & 9     & 4     & 10    & \cellcolor[rgb]{ .749,  .749,  .749}\textbf{7} & 11    & 5     & 9     & 7     & \cellcolor[rgb]{ .749,  .749,  .749}\textbf{8} & \cellcolor[rgb]{ .749,  .749,  .749}\textbf{6} & 8     & 10    & 8 \\
    \hline
    \texttt{UM-LR} & 5     & 3     & 6     & 7     & 4     & 3     & 6     & 12    & 4     & 9     & 5     & 10    & 7     & 10    & 5     & 4     & 4     & 3     & 5     & 7 \\
    \hline
    \texttt{CLIFE-NB} & 3     & 4     & 4     & 3     & 1     & 3     & 6     & 9     & 1     & 8     & 4     & 6     & 8     & 7     & 4     & 6     & 4     & 5     & 4     & 4 \\
    \hline
    \texttt{CLIFE-KNN} & 3     & 3     & \cellcolor[rgb]{ .749,  .749,  .749}\textbf{7} & 4     & 2     & 3     & 1     & 12    & 4     & 7     & 3     & 6     & 7     & 6     & 4     & 4     & 4     & 4     & 6     & 4 \\
    \hline
    \texttt{FeSCH-RF} & 3     & 3     & 4     & 2     & 3     & 3     & 5     & 10    & 1     & 6     & 2     & 8     & 8     & 4     & 4     & 5     & 4     & 5     & 4     & 6 \\
    \hline
    \texttt{GIS-NB} & 1     & 3     & 1     & 3     & 1     & 1     & 1     & 9     & 4     & 10    & 5     & 5     & 8     & 7     & 4     & 4     & 4     & 6     & 2     & 5 \\
    \hline
    \texttt{FeSCH-LR} & 4     & 3     & 4     & 1     & 2     & 3     & 5     & 2     & 1     & 7     & 3     & 7     & 5     & 5     & 4     & 4     & 5     & 7     & 7     & 6 \\
    \hline
    \texttt{CLIFE-SVM} & 4     & 3     & 4     & 4     & 3     & 2     & 1     & 7     & 4     & 2     & 3     & 8     & 6     & 6     & 4     & 7     & 3     & 3     & 8     & 3 \\
    \hline
    \texttt{TD-RF} & 3     & 3     & 6     & 3     & 2     & 3     & 5     & 6     & 3     & 5     & 3     & 2     & 8     & 3     & 4     & 4     & 3     & 5     & 6     & 3 \\
    \hline
    \texttt{TD-LR} & 2     & 1     & 4     & 3     & 2     & 3     & 5     & 9     & 1     & 4     & 3     & 6     & 8     & 6     & 4     & 4     & 3     & 6     & 9     & 3 \\
    \hline
    \texttt{TD-MLP} & 4     & 3     & 4     & 3     & 1     & 3     & 1     & 7     & 4     & 7     & 3     & 8     & 7     & 4     & 2     & 5     & 3     & 5     & 9     & 3 \\
    \hline
    \texttt{TD-DT} & 4     & 6     & 6     & 3     & 2     & 3     & 4     & 8     & 1     & 6     & 3     & 7     & 6     & 5     & 1     & 3     & 3     & 5     & 2     & 3 \\
    \hline
    \texttt{FeSCH-DT} & 4     & 3     & 3     & 3     & 2     & 3     & 5     & 8     & 2     & 7     & 3     & 5     & 2     & 8     & 1     & 2     & 3     & 2     & 5     & 3 \\
    \hline
    \texttt{VCB-SVM} & 4     & 2     & \cellcolor[rgb]{ .749,  .749,  .749}\textbf{7} & 3     & 2     & 3     & 2     & 2     & 1     & 2     & 2     & 2     & 5     & 3     & 4     & 2     & 1     & 1     & 1     & 2 \\
    \hline
    \texttt{CDE\_SMOTE-RF} & 3     & 4     & 4     & 5     & 2     & 3     & 1     & 1     & 4     & 1     & 1     & 1     & 1     & 3     & 2     & 1     & 1     & 1     & 1     & 1 \\
    \hline
    \texttt{CDE\_SMOTE-KNN} & 4     & 3     & 4     & 3     & 1     & 3     & 5     & 1     & 4     & 1     & 1     & 1     & 1     & 4     & 6     & 1     & 3     & 2     & \cellcolor[rgb]{ .749,  .749,  .749}\textbf{12} & 1 \\
    \hline
    \texttt{FSS\_bagging-RF} & 2     & \cellcolor[rgb]{ .749,  .749,  .749}\textbf{8} & 5     & 3     & 1     & 3     & 1     & 4     & 4     & 3     & 2     & 2     & 6     & 1     & 4     & 3     & 3     & 6     & 2     & 2 \\
    \hline
    \texttt{FSS\_bagging-NB} & 2     & 3     & 2     & 4     & 3     & 2     & 3     & 11    & 4     & 6     & 3     & 3     & 8     & 1     & 2     & 2     & 3     & 2     & 3     & 3 \\
    \hline
    \texttt{FSS\_bagging-LR} & 3     & 3     & 5     & 3     & 3     & 3     & 6     & 5     & 1     & 4     & 2     & 8     & 9     & 2     & 3     & 4     & 4     & 4     & 2     & 5 \\
    \hline
    \texttt{HISNN-NB} & 4     & 1     & 1     & 1     & 1     & 1     & 6     & 3     & 4     & 4     & 3     & 4     & 4     & 3     & 2     & 2     & 3     & 3     & 2     & 2 \\
    \hline
    \texttt{BiLO-CPDP} & \cellcolor[rgb]{ .749,  .749,  .749}\textbf{6} & \cellcolor[rgb]{ .749,  .749,  .749}\textbf{8} & \cellcolor[rgb]{ .749,  .749,  .749}\textbf{7} & 6     & \cellcolor[rgb]{ .749,  .749,  .749}\textbf{5} & \cellcolor[rgb]{ .749,  .749,  .749}\textbf{4} & 7     & 13    & \cellcolor[rgb]{ .749,  .749,  .749}\textbf{5} & \cellcolor[rgb]{ .749,  .749,  .749}\textbf{11} & 6     & \cellcolor[rgb]{ .749,  .749,  .749}\textbf{12} & \cellcolor[rgb]{ .749,  .749,  .749}\textbf{10} & \cellcolor[rgb]{ .749,  .749,  .749}\textbf{11} & 7     & \cellcolor[rgb]{ .749,  .749,  .749}\textbf{8} & \cellcolor[rgb]{ .749,  .749,  .749}\textbf{6} & \cellcolor[rgb]{ .749,  .749,  .749}\textbf{9} & 11    & \cellcolor[rgb]{ .749,  .749,  .749}\textbf{9} \\
    \hline
    \end{tabular}
    \centering
 \begin{tablenotes}
    \footnotesize
    \item The raw AUC values can be found in our repository: \url{https://github.com/COLA-Laboratory/ase2020}
    \end{tablenotes}
\end{table*}

\subsection{Experimental Procedure}
\label{sec:procedure}

Our experimental procedure follows the three-phases workflow of \texttt{BiLO-CPDP} introduced in~\pref{sec:architecture}. Here we explain the corresponding settings for each phase.
\begin{itemize}
    \item In the \textit{data pre-processing} phase for all peer CPDP techniques, all projects in this work will be used as target domain data in a round-robin manner, forming 20 different CPDP tasks. This aims to mitigate the potential bias in conclusion.
    
    \item In the \textit{optimization} phase, each CPDP task is allocated with an overall time budget of one hour (i.e., $3,600$ seconds, as suggested by Feurer et al.~\cite{FeurerKESBH15}) while setting each lower-level exploitation as 20 seconds in \texttt{BiLO-CPDP}. When applicable, the same budget is given to other state-of-the-art peer CPDP techniques that permit hyper-parameter optimization in the comparison, e.g., \texttt{Auto-sklearn}~\cite{FeurerKESBH15}. We apply the TPE algorithm implementation integrated in \texttt{Hyperopt}\footnote{\url{http://hyperopt.github.io/hyperopt/}}, a popular Python library for hyper-parameter tuning in machine learning~\cite{BergstraYC13}, for the lower-level routine of \texttt{BiLO-CPDP}.
    
    \item In the \textit{performance validation} phase, AUC is used as the performance metric. Due to the stochastic nature of \texttt{BiLO-CPDP} and some peer CPDP techniques considered, each technique is independently repeated 30 times for a given CPDP task and the mean AUC values are recorded for comparison.
\end{itemize}

\subsection{Ranking, Statistical Test and Effect Size}
\label{sec:statistics}

In our experiments, we use the following three statistical measures to interpret the statistical significance of our comparative results.

\begin{itemize}
    \item\textbf{Scott-Knott test:} Instead of merely comparing the raw AUC values, we apply the Scott-Knott test to rank the performance of different peer techniques over 30 runs on each project, as recommended by Mittas and Angelis~\cite{MittasA13}. In a nutshell, the Scott-Knott test uses a statistical test and effect size to divide the performance of peer techniques into several clusters. In particular, the performance of peer techniques within the same cluster are statistically insignificant, i.e., their overall AUC values are statistically equivalent. Note that the clustering process terminates until no split can be made. Finally, each cluster can be assigned a rank according to the mean AUC values achieved by the peer techniques within the cluster. In particular, since a greater AUC is preferred, the larger the rank is, the better performance of the technique achieves.
    
    \item\textbf{Wilcoxon signed-rank test:} We apply the Wilcoxon signed-rank test ~\cite{Wilcoxon1945IndividualCB} with a significant level $p=0.05$~\cite{ArcuriB11} to investigate the statistical significance of the comparisons. It is a non-parametric statistical test that makes little assumption about the underlying distribution of the data and has been recommended in software engineering research~\cite{ArcuriB11}.
    
    \item\textbf{$\mathbf{A_{12}}$ effect size:} To ensure the resulted differences are not generated from a trivial effect, we apply $A_{12}$~\cite{Vargha2000ACA} as the effect size measure to evaluate the probability that one technique is better than another. According to Vargha and Delaney~\cite{Vargha2000ACA}, when comparing \texttt{BiLO-CDPD} with another peer technique in our experiments, $A_{12}=0.5$ means they are equivalent. $A_{12}>0.5$ denotes that \texttt{BiLO-CDPD} is better for more than 50\% of the times. In particular, $0.56\leq A_{12}<0.64$ indicates a small effect size while $0.64 \leq A_{12} < 0.71$ and $A_{12} \geq 0.71$ mean a medium and a large effect size, respectively. 
\end{itemize}
Note that both Wilcoxon signed-rank test and $A_{12}$ are also used in the Scott-Knott test for generating the clusters.

\subsection{Research Questions}
\label{sec:RQs}

We seek to answer the following four research questions (RQs) through our experimental evaluation:

\begin{itemize}
    \item\textbf{\underline{RQ1:}} Is \texttt{BiLO-CPDP} able to automatically configure a CPDP model having better performance than the existing CPDP techniques under their reported settings?
    \item\textbf{\underline{RQ2:}} How is the performance of \texttt{BiLO-CPDP} comparing with \texttt{Auto-} \texttt{Sklearn}, a state-of-the-art AutoML tool?
    \item\textbf{\underline{RQ3:}} Is the bi-level programming in \texttt{BiLO-CPDP} beneficial?
    \item\textbf{\underline{RQ4:}} Given a limited computational budget, which level in \texttt{BiLO-CPDP} is more important and deserves more budget?
\end{itemize}


\section{Results and Discussions}
\label{sec:result}

In this section, we present and discuss the results of our empirical experiments and address the RQs posed in~\pref{sec:RQs}.


\subsection{Comparison with Existing CPDP Work}
\label{sec:rq1}

\begin{figure}[t!]
    \centering
    \includegraphics[width=\linewidth]{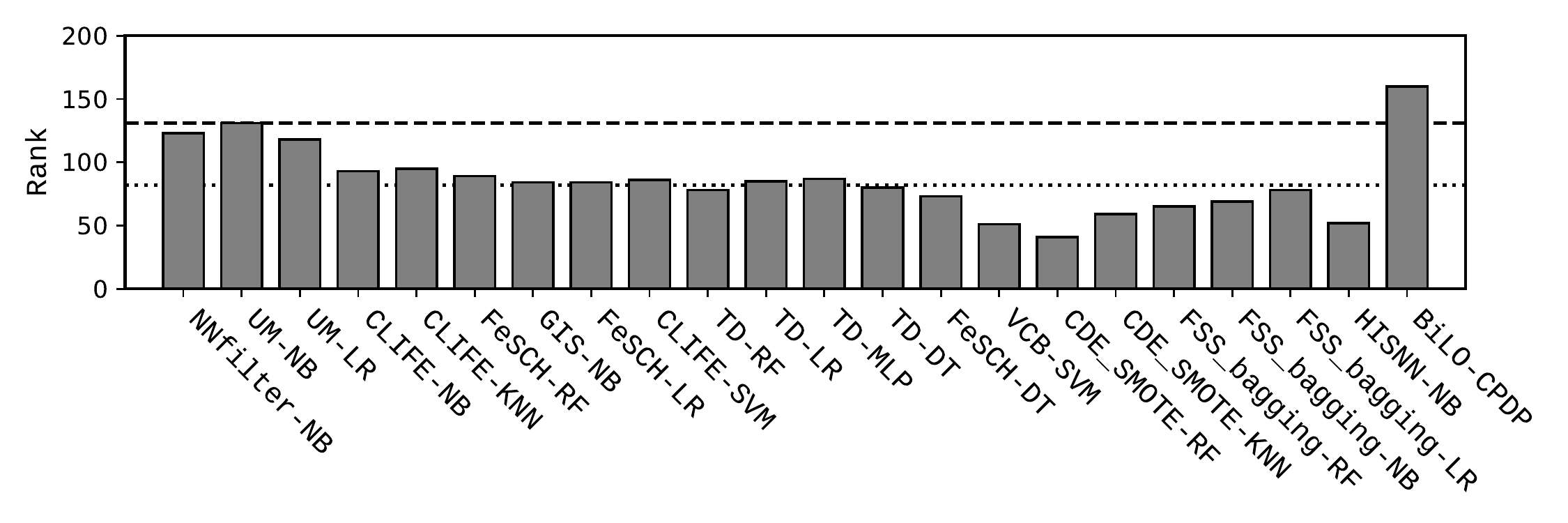}
    \caption{Total ranks achieved by \texttt{BiLO-CPDP} (the right most one) and the 21 peer techniques (the larger rank, the better; the dashed line and dotted line denote the best and average result over the 21 peer techniques, respectively).}
    \label{fig:rq1_rank}
\end{figure}

\subsubsection{Method}
In order to answer RQ1, we use the transfer learners and classifiers collected in Tables~\ref{tab:transfer_learner} and~\ref{tab:classifier} to constitute 21 peer CPDP techniques in comparison with \texttt{BiLO-CPDP}. Note that although there are only 13 transfer learners listed in~\pref{tab:transfer_learner}, some of them are combined with more than one classifier to constitute different CPDP models used in the literature (e.g., \texttt{TD} is combined with classifiers \texttt{RF}, \texttt{LR}, \texttt{MLP} and \texttt{DT} that constitute four different CPDP models in~\cite{herbold2013training}). For the parameter settings, we use the tuned values as reported in the corresponding work. 

\subsubsection{Results and Analysis}
From the experimental results on the Scott-Knott test shown in~\pref{tab:rq1}, it is clear to see that \texttt{BiLO-CPDP} is the best on 14 out of 20 (70\%) projects, second only to one other on five cases. In contrast, most of the other peer CPDP techniques, albeit hand crafted by domain experts, are not as competitive as \texttt{BiLO-CPDP}. In particular, \texttt{NNfilter-NB} is the most outstanding peer technique that is the best on only 7 out 20 (35\%) projects while the other peer techniques rarely take the best rank across all 20 projects. Noteworthily, the performance of \texttt{NNfilter-NB} ties with \texttt{BiLO-CPDP} in two of its best results. In terms of the total ranks achieved over all projects, as shown in~\pref{fig:rq1_rank}, we can observe the clear superiority of \texttt{BiLO-CPDP} which is at least 50\% better than the other 21 peer techniques. Furthermore, we notice that the superior performance of \texttt{BiLO-CPDP} is consistent across all 20 projects in view of its top three ranked positions achieved in all projects. In contrast, the performance of existing CPDP techniques exhibit clear variations depending on the underlying target projects.


\vspace{0.5em}
\noindent
\framebox{\parbox{\dimexpr\linewidth-2\fboxsep-2\fboxrule}{
	\textbf{\underline{Response to RQ1:}} \textit{\texttt{BiLO-CPDP} is generally better than the other 21 existing CPDP techniques over all 20 projects. Unlike others that were hand-crafted by domain experts to certain extents,  \texttt{BiLO-CPDP} builds an effective CPDP model in a completely automated manner, leading to highly competitive performance over different projects. }
}}

\begin{table}[t!]
    \small
    \setlength{\tabcolsep}{0.8mm}
    \centering
    \caption{Mean AUC (standard deviation) for \texttt{BiLO-CPDP} and \texttt{Auto-Sklearn} over 30 runs (gray=better; bold=$p$<.05).}
	\label{tab:rq2}%
    \begin{tabular}{c|c|c|c}
    \hline
          \textbf{Project} & \textbf{BiLO-CPDP} & \textbf{Auto-sklearn} & \textbf{$p$-value} \\
    \hline
    \texttt{poi}   & \cellcolor[rgb]{ .749,  .749,  .749}\textbf{8.1703E-1 (4.38E-3)} & 6.5262E-1 (6.98E-3) & 1.71E-6 \\
    \hline
    \texttt{synapse} & \cellcolor[rgb]{ .749,  .749,  .749}\textbf{7.1999E-1 (7.72E-3)} & 6.0183E-1 (3.68E-3) & 1.65E-6 \\
    \hline
    \texttt{Zxing} & \cellcolor[rgb]{ .749,  .749,  .749}\textbf{6.3949E-1 (5.37E-3)} & 6.2615E-1 (1.45E-6) & 1.91E-6 \\
    \hline
    \texttt{ant}   & \cellcolor[rgb]{ .749,  .749,  .749}\textbf{8.0006E-1 (8.26E-3)} & 7.4530E-1 (7.63E-3) & 1.71E-6 \\
    \hline
    \texttt{log4j} & \cellcolor[rgb]{ .749,  .749,  .749}\textbf{8.4196E-1 (1.52E-2)} & 6.0965E-1 (1.19E-2) & 1.57E-6 \\
    \hline
    \texttt{Safe}  & \cellcolor[rgb]{ .749,  .749,  .749}\textbf{7.9923E-1 (2.09E-2)} & 6.7513E-1 (6.15E-3) & 1.64E-6 \\
    \hline
    \texttt{ivy}   & \cellcolor[rgb]{ .749,  .749,  .749}\textbf{8.0657E-1 (3.51E-3)} & 7.2407E-1 (9.64E-4) & 1.19E-6 \\
    \hline
    \texttt{PDE}   & \cellcolor[rgb]{ .749,  .749,  .749}\textbf{6.8539E-1 (2.57E-3)} & 5.9781E-1 (2.22E-16) & 1.62E-6 \\
    \hline
    \texttt{camel} & \cellcolor[rgb]{ .749,  .749,  .749}\textbf{6.2228E-1 (4.01E-3)} & 5.9006E-1 (1.11E-16) & 1.62E-6 \\
    \hline
    \texttt{lucene} & \cellcolor[rgb]{ .749,  .749,  .749}\textbf{7.1065E-1 (8.13E-3)} & 6.4408E-1 (4.87E-6) & 1.37E-6 \\
    \hline
    \texttt{JDT}   & \cellcolor[rgb]{ .749,  .749,  .749}\textbf{7.3705E-1 (1.09E-2)} & 6.7517E-1 (1.11E-16) & 1.66E-6 \\
    \hline
    \texttt{jEdit} & \cellcolor[rgb]{ .749,  .749,  .749}\textbf{8.5207E-1 (3.77E-2)} & 7.1589E-1 (5.70E-3) & 1.68E-6 \\
    \hline
    \texttt{EQ}    & \cellcolor[rgb]{ .749,  .749,  .749}\textbf{7.1714E-1 (1.34E-2)} & 6.0201E-1 (3.33E-16) & 1.73E-6 \\
    \hline
    \texttt{velocity} & \cellcolor[rgb]{ .749,  .749,  .749}\textbf{7.0220E-1 (8.40E-3)} & 6.0896E-1 (4.50E-2) & 1.61E-6 \\
    \hline
    \texttt{Tomcat} & \cellcolor[rgb]{ .749,  .749,  .749}\textbf{7.7295E-1 (1.40E-3)} & 7.3892E-1 (1.32E-2) & 1.45E-7 \\
    \hline
    \texttt{Apache} & \cellcolor[rgb]{ .749,  .749,  .749}7.4808E-1 (8.27E-3) & 7.4787E-1 (3.33E-16) & 6.58E-1 \\
    \hline
    \texttt{ML}    & \cellcolor[rgb]{ .749,  .749,  .749}\textbf{6.4966E-1 (1.58E-3)} & 6.1708E-1 (2.22E-16) & 1.73E-6 \\
    \hline
    \texttt{xerces} & \cellcolor[rgb]{ .749,  .749,  .749}\textbf{7.1552E-1 (1.03E-2)} & 5.9892E-1 (6.03E-3) & 1.71E-6 \\
    \hline
    \texttt{LC}    & \cellcolor[rgb]{ .749,  .749,  .749}\textbf{7.0859E-1 (1.89E-2)} & 6.2476E-1 (1.11E-16) & 1.73E-6 \\
    \hline
    \texttt{xalan} & \cellcolor[rgb]{ .749,  .749,  .749}\textbf{7.6250E-1 (7.67E-3)} & 6.7732E-1 (2.31E-2) & 1.71E-6 \\
    \hline
    \end{tabular}%
\end{table}

\begin{figure}[t!]
    \centering
    \includegraphics[width=\linewidth]{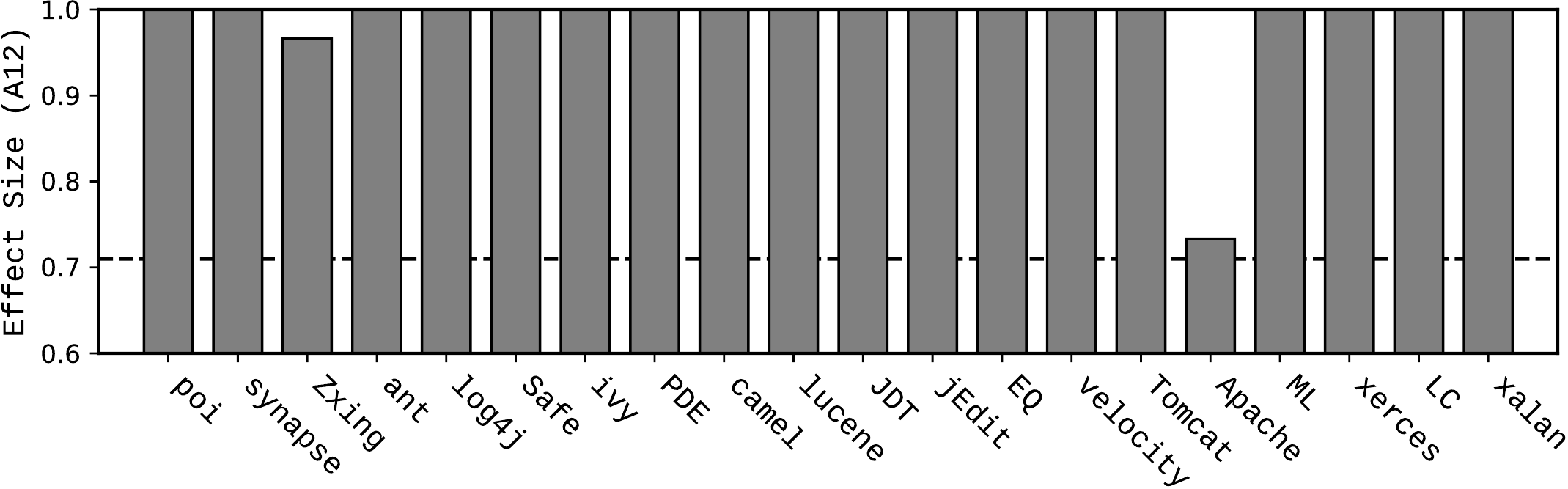}
    \caption{$A_{12}$ result between \texttt{BiLO-CPDP} and \texttt{Auto-Sklearn} over 30 runs ($A_{12}>0.5$ means \texttt{BiLO-CPDP} is better).}
    \label{fig:bi-au}
\end{figure}

\subsection{Comparison with \texttt{Auto-Sklearn}}
\label{sec:rq2}

\subsubsection{Method}

In principle, \texttt{BiLO-CPDP} is an AutoML tool that automatically searches for the right combination of transfer learner and classifier and their optimized hyper-parameter settings for a given CPDP task. To validate its competitiveness from the perspective of AutoML, we compare the performance of \texttt{BiLO-CPDP} with \texttt{Auto-Sklearn}\footnote{\url{https://automl.github.io/auto-sklearn/master/}}~\cite{FeurerKESBH15}, a state-of-the-art and readily available AutoML tool that can also optimize the combination and its parameters. 


\subsubsection{Results and Analysis}

From the comparison results of AUC values shown in~\pref{tab:rq2}, we clearly see the overwhelmingly superior performance of \texttt{BiLO-CPDP} versus \texttt{Auto-Sklearn} where the AUC values obtained by \texttt{BiLO-CPDP} are all better than those of \texttt{Auto-Sklearn}. In particular, all those better results, except on \texttt{Apache}, are statistically significant, according to the $p$ values shown in the last column of~\pref{tab:rq2}. Furthermore, as shown in~\pref{fig:bi-au}, all $A_{12}$ values suggest a large effect size. In particular, we see an overwhelming $A_{12}=1$, except only on \texttt{Zxing} and \texttt{Apache}. These indicate that the improvements on the AUC results brought by \texttt{BiLO-CPDP} over that of the \texttt{Auto-Sklearn} are significantly large in general.

The results are caused by the fact that \texttt{Auto-Sklearn} does not have a bi-level structure, hence it encodes all transfer learner and classifier combinations along with their corresponding hyper-para- meter settings into an integrated solution at a single-level, which is solved by the SMAC algorithm~\cite{DBLP:conf/lion/HutterHL11}. During its optimization process, a combination of transfer learner and classifier is selected first. Thereafter, the variables corresponding to the hyper-parameters of the chosen transfer learner and the classifier become active while the remaining variables are set to be dummy. By this means, the total number of variables considered in \texttt{Auto-Sklean} goes up to 93, resulting a unnecessarily much larger search space comparing with \texttt{BiLO-CPDP}. Given the limited budget, \texttt{Auto-Sklearn} therefore ends up with a less effective exploration of both useful combinations of transfer learners and classifiers and their hyper-parameter settings.

\vspace{0.5em}
\noindent
\framebox{\parbox{\dimexpr\linewidth-2\fboxsep-2\fboxrule}{
    \textbf{\underline{Response to RQ2:}} \textit{Comparing with the state-of-the-art AutoML tool \texttt{Auto-Sklearn}, \texttt{BiLO-CPDP} achieves significantly better results given a limited computational budget.}

}}

\subsection{Comparison with Single-Level Variant}
\label{sec:rq3}



\subsubsection{Method}

It is conservative to curious about the usefulness brought by this bi-level programming formulation and why not simply formulating a single-level problem that consists of both combination and parameters. The comparison with \texttt{Auto-Sklearn}, which is at a single-level, partially validates this concern, but the results can be biased by the fact that it uses a different optimization algorithm. To fully evaluate the effectiveness of bi-level programming, we develop a single-level variant of \texttt{BiLO-CPDP}, dubbed \texttt{SLO-CPDP}, which differs from \texttt{BiLO-CPDP} only on the solution representation. 


Specifically, \texttt{SLO-CPDP} is similar to \texttt{Auto-Sklearn} in the sense that they both work on single-level optimization --- the transfer learner and classifier, together with their hyper-parameters, are encoded as a single solution representation. However, the difference is that \texttt{SLO-CPDP} exploits the \texttt{TPE} algorithm as the Bayesian optimizer, which is identical to \texttt{BiLO-CPDP}. \texttt{Auto-Sklearn}, in contrast, uses the classic SMAC algorithm that leverages Random Forest to build the surrogate model.




\subsubsection{Results and Analysis}

\begin{figure}[t!]
    \centering
    \includegraphics[width=\linewidth]{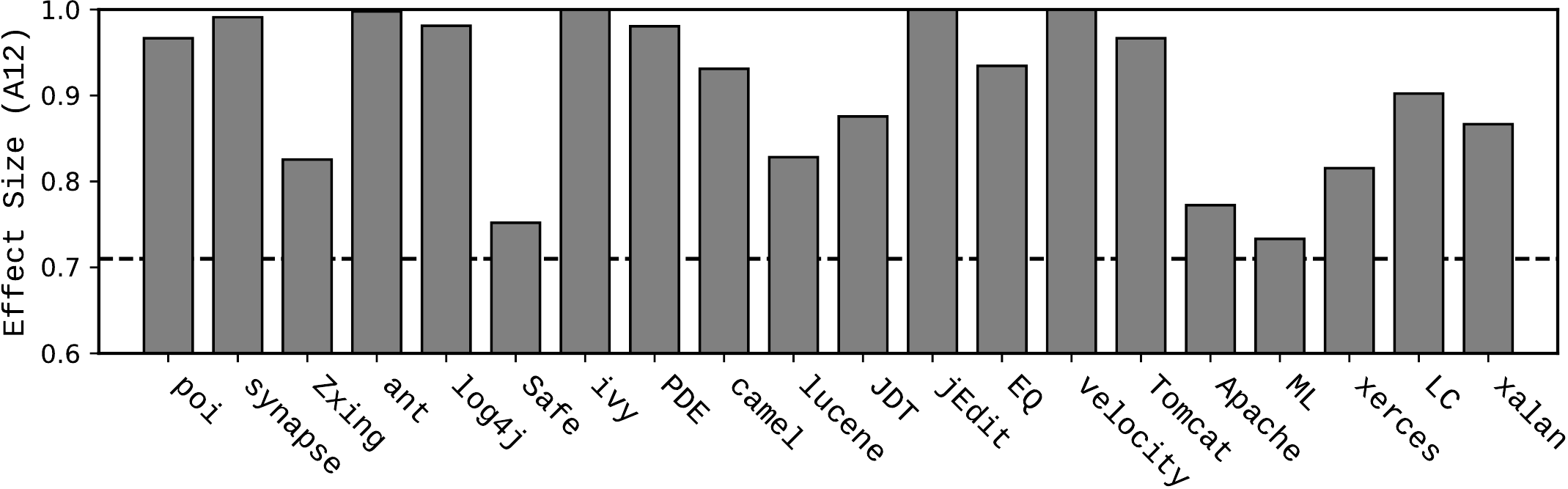}
    \caption{$A_{12}$ result between \texttt{BiLO-CPDP} and \texttt{SLO-CPDP} over 30 runs ($A_{12}>0.5$ means \texttt{BiLO-CPDP} is better).}
    \label{fig:bi_si}
\end{figure}

From the AUC values shown in~\pref{tab:rq3}, we observe a rather superior performance achieved by \texttt{BiLO-CPDP} over \texttt{SLO-CPDP}. Specifically, \texttt{BiLO-CPDP} again obtains a better AUC value on all 20 projects. In particular, all better results are with statistical significance ($p$<.05), as shown in the last column of~\pref{tab:rq3}. Furthermore, from~\pref{fig:bi_si}, we find that the differences between the AUC values achieved by \texttt{BiLO-CPDP} and \texttt{SLO-CPDP} are with a large effect size. Given such a result, we can infer that the ineffectiveness of \texttt{SLO-CPDP} can be attributed to the enlarged search space caused by the unwise coupling of transfer learner and classifier combination along with their parameters at a single-level.

\begin{table}[t!]
    \small
    \setlength{\tabcolsep}{0.8mm}
    \centering
    \caption{Mean AUC (standard deviation) for \texttt{BiLO-CPDP} and \texttt{SLO-CPDP} over 30 runs (gray=better; bold=$p$<.05).}
	\label{tab:rq3}
    \begin{tabular}{c|c|c|c}
    \hline
          \textbf{Project} & \texttt{\textbf{BiLO-CPDP}} & \texttt{\textbf{SLO-CPDP}} & $p$-\textbf{value} \\
    \hline
    \texttt{poi}   & \cellcolor[rgb]{ .749,  .749,  .749}\textbf{8.1703E-1 (4.38E-3)} & 5.7493E-1 (2.71E-1) & 1.92E-6 \\
    \hline
    \texttt{synapse} & \cellcolor[rgb]{ .749,  .749,  .749}\textbf{7.1999E-1 (7.72E-3)} & 5.0601E-1 (2.33E-1) & 1.73E-6 \\
    \hline
    \texttt{Zxing} & \cellcolor[rgb]{ .749,  .749,  .749}\textbf{6.3949E-1 (5.37E-3)} & 5.4209E-1 (1.46E-1) & 1.92E-6 \\
    \hline
    \texttt{ant}   & \cellcolor[rgb]{ .749,  .749,  .749}\textbf{8.0006E-1 (8.26E-3)} & 6.3099E-1 (1.76E-1) & 1.73E-6 \\
    \hline
    \texttt{log4j} & \cellcolor[rgb]{ .749,  .749,  .749}\textbf{8.4196E-1 (1.52E-2)} & 6.2807E-1 (2.49E-1) & 2.60E-6 \\
    \hline
    \texttt{Safe}  & \cellcolor[rgb]{ .749,  .749,  .749}\textbf{7.9923E-1 (2.09E-2)} & 7.4254E-1 (3.96E-2) & 5.74E-5 \\
    \hline
    \texttt{ivy}   & \cellcolor[rgb]{ .749,  .749,  .749}\textbf{8.0657E-1 (3.51E-3)} & 6.3528E-1 (1.77E-1) & 1.73E-6 \\
    \hline
    \texttt{PDE}   & \cellcolor[rgb]{ .749,  .749,  .749}\textbf{6.8539E-1 (2.57E-3)} & 5.0874E-1 (2.52E-1) & 1.73E-6 \\
    \hline
    \texttt{camel} & \cellcolor[rgb]{ .749,  .749,  .749}\textbf{6.2228E-1 (4.01E-3)} & 4.7330E-1 (2.15E-1) & 5.22E-6 \\
    \hline
    \texttt{lucene} & \cellcolor[rgb]{ .749,  .749,  .749}\textbf{7.1065E-1 (8.13E-3)} & 5.8568E-1 (2.05E-1) & 1.15E-4 \\
    \hline
    \texttt{JDT}   & \cellcolor[rgb]{ .749,  .749,  .749}\textbf{7.3705E-1 (1.09E-2)} & 6.2603E-1 (1.81E-1) & 1.36E-5 \\
    \hline
    \texttt{jEdit} & \cellcolor[rgb]{ .749,  .749,  .749}\textbf{8.5207E-1 (3.77E-2)} & 5.5203E-1 (2.50E-1) & 1.73E-6 \\
    \hline
    \texttt{EQ}    & \cellcolor[rgb]{ .749,  .749,  .749}\textbf{7.1714E-1 (1.34E-2)} & 5.4016E-1 (2.21E-1) & 2.88E-6 \\
    \hline
    \texttt{velocity} & \cellcolor[rgb]{ .749,  .749,  .749}\textbf{7.0220E-1 (8.40E-3)} & 5.3123E-1 (2.11E-1) & 1.73E-6 \\
    \hline
    \texttt{Tomcat} & \cellcolor[rgb]{ .749,  .749,  .749}\textbf{7.7295E-1 (1.40E-3)} & 5.6400E-1 (2.40E-1) & 1.92E-6 \\
    \hline
    \texttt{Apache} & \cellcolor[rgb]{ .749,  .749,  .749}\textbf{7.4808E-1 (8.27E-3)} & 5.2924E-1 (1.23E-1) & 1.01E-6 \\
    \hline
    \texttt{ML}    & \cellcolor[rgb]{ .749,  .749,  .749}\textbf{6.4966E-1 (1.58E-3)} & 5.8287E-1 (1.53E-1) & 7.16E-4 \\
    \hline
    \texttt{xerces} & \cellcolor[rgb]{ .749,  .749,  .749}\textbf{7.1552E-1 (1.03E-2)} & 6.3384E-1 (1.29E-1) & 6.87E-5 \\
    \hline
    \texttt{LC}    & \cellcolor[rgb]{ .749,  .749,  .749}\textbf{7.0859E-1 (1.89E-2)} & 5.3199E-1 (2.38E-1) & 1.02E-5 \\
    \hline
    \texttt{xalan} & \cellcolor[rgb]{ .749,  .749,  .749}\textbf{7.6250E-1 (7.67E-3)} & 5.9866E-1 (2.21E-1) & 5.79E-5 \\
    \hline
    \end{tabular}%
   
\end{table}%

\vspace{0.5em}
\noindent
\framebox{\parbox{\dimexpr\linewidth-2\fboxsep-2\fboxrule}{
    \textbf{\underline{Response to RQ3:}} \textit{The bi-level programming in \texttt{BiLO-CPDP} considerably contributes to its effectiveness. In contrast to the single-level where the combination and parameters are formulated in a ``flat" way, bi-level programming significantly reduces the search space and steer the search in a hierarchical manner, leading to better performance under a limited budget.}
}}

\subsection{Impact of Budget for the Two Levels}
\label{sec:two_level_importance}


\subsubsection{Method}

In practice, it is not uncommon that the resource for defect prediction, particular the time budget, is limited. In our experiments, the total budget allocated to \texttt{BiLO-CPDP} is one hour in total, following the best practice in the AutoML community~\cite{FeurerKESBH15}. However, the unique bi-level programming formulated in \texttt{BiLO-CPDP} allows us a flexible control over the budget allocated to the two levels, hence it is interested to know how their budget allocations may impact the performance. To this end, within the one hour total budget, we set two budget allocation strategies: one with high budget to the upper-level, dubbed \texttt{BiLO-CPDP(h)}, that allows 20 seconds for each low-level optimization, leaving more resources for exploring the combinations at the upper-level. Note that 20 seconds are very short for some model training thus is counter-intuitive. This is also the default setting in \texttt{BiLO-CPDP} we used for other experiments. Another one preserves high budget to the low-level, dubbed \texttt{BiLO-CPDP(l)}, in which the lower-level optimization is allocated with 100 training and AUC evaluations. This allows a low-level routine to consume at least 80 seconds (the smallest amount of time required to complete 100 evaluations among all combinations) for more sufficient exploration of the hyper-parameters.

\begin{table}[t!]
    \small
    \setlength{\tabcolsep}{0.8mm}
    \centering
    \caption{Mean AUC (standard deviation) for \texttt{BiLO-CPDP(h)} and \texttt{BiLO-CPDP(l)} over 30 runs (gray=better; bold=$p$<.05).}
	\label{tab:rq4}
    \begin{tabular}{c|c|c|c}
    \hline
          \textbf{Project} & \textbf{BiLO-CPDP(h)} & \textbf{BiLO-CPDP(l)} & \textbf{$p$-value} \\
    \hline
    \texttt{poi}   & \cellcolor[rgb]{ .749,  .749,  .749}\textbf{8.1703E-1 (4.38E-3)} & 5.9447E-1 (2.74E-1) & 3.85E-6 \\
    \hline
    \texttt{synapse} & \cellcolor[rgb]{ .749,  .749,  .749}\textbf{7.1999E-1 (7.72E-3)} & 6.1897E-1 (1.26E-1) & 1.73E-6 \\
    \hline
    \texttt{Zxing} & \cellcolor[rgb]{ .749,  .749,  .749}\textbf{6.3949E-1 (5.37E-3)} & 6.1108E-1 (2.10E-2) & 8.46E-6 \\
    \hline
    \texttt{ant}   & \cellcolor[rgb]{ .749,  .749,  .749}\textbf{8.0006E-1 (8.26E-3)} & 4.6150E-1 (3.31E-1) & 1.91E-6 \\
    \hline
    \texttt{log4j} & \cellcolor[rgb]{ .749,  .749,  .749}\textbf{8.4196E-1 (1.52E-2)} & 5.1593E-1 (2.91E-1) & 2.46E-6 \\
    \hline
    \texttt{Safe}  & \cellcolor[rgb]{ .749,  .749,  .749}\textbf{7.9923E-1 (2.09E-2)} & 7.1771E-1 (1.38E-1) & 2.87E-6 \\
    \hline
    \texttt{ivy}   & \cellcolor[rgb]{ .749,  .749,  .749}\textbf{8.0657E-1 (3.51E-3)} & 5.9659E-1 (3.04E-1) & 5.23E-6 \\
    \hline
    \texttt{PDE}   & \cellcolor[rgb]{ .749,  .749,  .749}\textbf{6.8539E-1 (2.57E-3)} & 4.3027E-1 (3.05E-1) & 1.71E-6 \\
    \hline
    \texttt{camel} & \cellcolor[rgb]{ .749,  .749,  .749}\textbf{6.2228E-1 (4.01E-3)} & 3.9054E-1 (2.79E-1) & 7.90E-6 \\
    \hline
    \texttt{lucene} & \cellcolor[rgb]{ .749,  .749,  .749}\textbf{7.1065E-1 (8.13E-3)} & 5.3341E-1 (2.69E-1) & 1.12E-5 \\
    \hline
    \texttt{JDT}   & \cellcolor[rgb]{ .749,  .749,  .749}\textbf{7.3705E-1 (1.09E-2)} & 3.7420E-1 (3.51E-1) & 3.85E-6 \\
    \hline
    \texttt{jEdit} & \cellcolor[rgb]{ .749,  .749,  .749}\textbf{8.5207E-1 (3.77E-2)} & 5.0339E-1 (3.30E-1) & 1.72E-6 \\
    \hline
    \texttt{EQ}    & \cellcolor[rgb]{ .749,  .749,  .749}\textbf{7.1714E-1 (1.34E-2)} & 4.2035E-1 (3.22E-1) & 8.37E-6 \\
    \hline
    \texttt{velocity} & \cellcolor[rgb]{ .749,  .749,  .749}\textbf{7.0220E-1 (8.40E-3)} & 5.0660E-1 (2.55E-1) & 1.92E-6 \\
    \hline
    \texttt{Tomcat} & \cellcolor[rgb]{ .749,  .749,  .749}\textbf{7.7295E-1 (1.40E-3)} & 5.6762E-1 (2.87E-1) & 2.50E-6 \\
    \hline
    \texttt{Apache} & \cellcolor[rgb]{ .749,  .749,  .749}\textbf{7.4808E-1 (8.27E-3)} & 7.1257E-1 (2.62E-2) & 7.22E-6 \\
    \hline
    \texttt{ML }   & \cellcolor[rgb]{ .749,  .749,  .749}\textbf{6.4966E-1 (1.58E-3)} & 3.7510E-1 (3.07E-1) & 5.70E-6 \\
    \hline
    \texttt{xerces} & \cellcolor[rgb]{ .749,  .749,  .749}\textbf{7.1552E-1 (1.03E-2)} & 4.8507E-1 (2.72E-1) & 3.87E-6 \\
    \hline
    \texttt{LC}    & \cellcolor[rgb]{ .749,  .749,  .749}\textbf{7.0859E-1 (1.89E-2)} & 4.9327E-1 (3.00E-1) & 1.23E-4 \\
    \hline
    \texttt{xalan} & \cellcolor[rgb]{ .749,  .749,  .749}\textbf{7.6250E-1 (7.67E-3)} & 4.9144E-1 (3.05E-1) & 1.91E-6 \\
    \hline
    \end{tabular}%
\end{table}

\begin{figure}[t!]
    \centering
    \includegraphics[width=\linewidth]{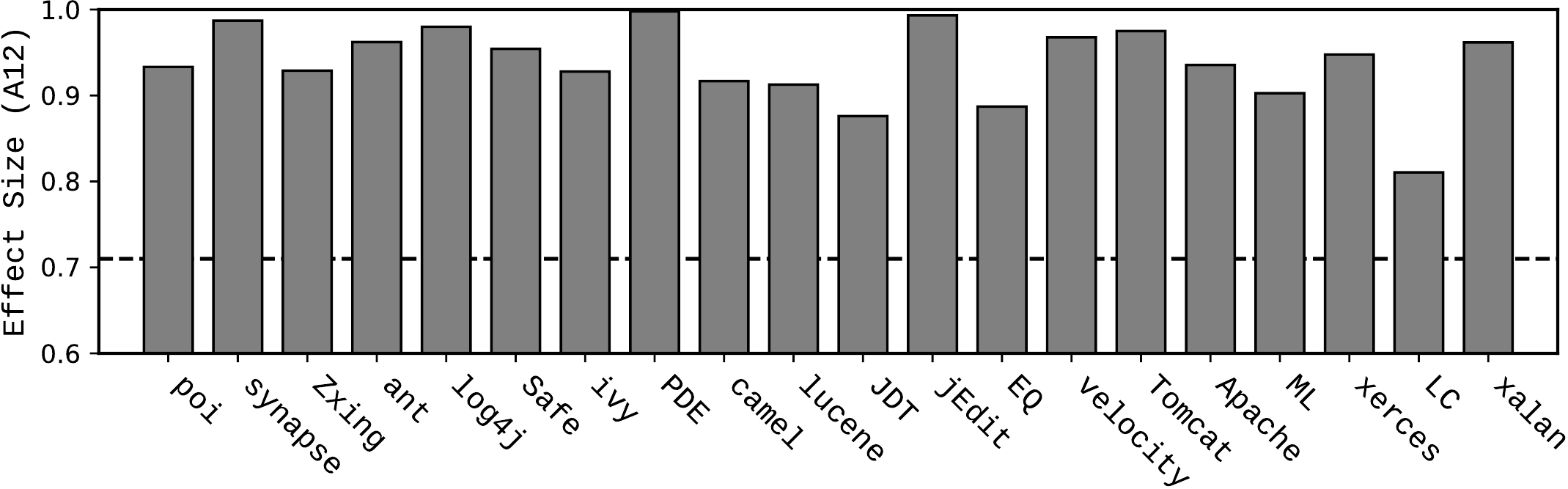}
    \caption{$A_{12}$ result between \texttt{BiLO-CPDP(h)} and \texttt{BiLO-CPDP(l)} over 30 runs ($A_{12}>0.5$ means \texttt{BiLO-CPDP(h)} is better).}
    \label{fig:time}
\end{figure}

\subsubsection{Results and Analysis}

As shown in~\pref{tab:rq4}, we can see that \texttt{BiLO-CPDP(h)} is overwhelmingly superior to \texttt{BiLO-} \texttt{CPDP(l)} where it obtains better AUC values on all 20 projects. In addition, from the comparison results of $A_{12}$ shown in~\pref{fig:time}, we can see that the differences between AUC values achieved by the two budget allocation strategies are categorized to have a large effect size.

The performance differences are due to the fact that the CPDP model training can be rather time-consuming and unfavorable, especially given a limited budget. Therefore, for \texttt{BiLO-CPDP(l)}, once a combination of transfer learner and classifier is selected at the upper-level routine, its initiative to favor better exploration of the hyper-parameters at the lower-level routine can easily consume a significant amount of the budget (the median computational time is around 300 seconds according to our offline statistics). This has caused the combinatorial space of transfer learners and classifiers to become severely under-explored. In contrast, by strictly restricting the budget at the lower-level optimization routine, \texttt{BiLO-CPDP(h)} suffers from a limited exploration of the hyper-parameter space, but permitting a sufficient chance to explore many combinations of transfer learners and classifiers. From the results, it evidences that exploring the combination space is more important than using the hyper-parameter space under a limited budget.

\vspace{0.5em}
\noindent
\framebox{\parbox{\dimexpr\linewidth-2\fboxsep-2\fboxrule}{
    \textbf{\underline{Response to RQ4:}} \textit{Given a limited budget, it is recommended to allocate more expenditure to the upper-level optimization routine in \texttt{BiLO-CPDP}. By this means, more combinations of transfer learner and classifier can be investigated even without fully optimized hyper-parameters, which is more beneficial to performance.}
}}

\section{Related Work}
\label{sec:related}

In the past decades, machine learning classifiers have become the core techniques for defect prediction, in which the success can be greatly affected by the setting of the classifiers' hyper-parameters~\cite{KoruL05}. This is a challenging issue, as Jiang et al.~\cite{JiangCM08} pointed out that simply using the default values are dreadful, causing severely bad performance of the prediction. The automated parameter optimization for defect predictors is therefore crucial. Indeed, a large scale empirical study by Tantithamthavorn et al.~\cite{Tantithamthavorn16,Tantithamthavorn19} found that well-tuned hyper-parameters can significantly boost the performance of the classifiers in defect prediction. Fu~\cite{fu2016tuning} even suggest that such optimization should become a standard practice in every single Software Engineering task. In light of this, Agrawal et al.~\cite{DBLP:conf/icse/AgrawalM18} have applied Differential Evolution to tune \texttt{SMOTE}, a pre-processor for handling data imbalance, for predicting software defects. Their work focus on within-project defect prediction though. Similarly, \texttt{DODGE}~\cite{8854183} is a recent tool that optimizes the parameters of data pre-processor and classifier. Although they aim for within-project case, the combination of pre-processor and classifier can be resemble to our CPDP task. However, their optimization assumes conservative hybridization of all the parameters and the combinations as a single-level optimization problem. 

The importance of automated parameter optimization remains stand in the context of CPDP, where the problem become even more complex as the parameters of transfer learners also come into play. Qu et al.~\cite{qu2018impact} have shown that the parameter settings of classifiers for CPDP are even more important. A few automated optimizers exist for CPDP, for example, Ozturk et al.~\cite{Ozturk19} and Qu et al.~\cite{QuCZJ18} examine various different optimization algorithms to tune CPDP models. Nevertheless, they focus only on the parameter tuning whilst ignore the combination of transfer learner and classifier during optimization. Indeed, Li et al.~\cite{LiXCWT20} further demonstrate that the parameter interactions between transfer learner and classifier, as well as their combination, also play an integral role to the prediction performance. \texttt{Auto-Sklearn}~\cite{FeurerKESBH15}, which is a widely-used generic tool to tune arbitrary machine learning algorithms, is also highly potential for CPDP tuning. However, again, its design has restricted that the combination of transfer learner and classifier along with their parameters need to be tuned as a single optimization problem, which worsen its performance compared with \texttt{BiLO-CPDP}, as we have shown in~\pref{sec:result}. 

Although the potentials of bi-level programming have been explored for other Software Engineering problems, e.g., code smell detection~\cite{sahin2014code} and test case generation~\cite{sahin2015model}, to the best of our knowledge, its adoption has never been reported in the context of CPDP. Our work is therefore unique to all aforementioned techniques in the sense that:

\begin{itemize}
    \item \texttt{BiLO-CPDP} is the first of its kind to formulate bi-level programming for the parameter optimization of CPDP.
    \item \texttt{BiLO-CPDP} automatically optimizes not only the parameters, but also discover the possible combination of transfer learner and classifier from a given portfolio. 
    \item We show that exploring the combination of transfer learner and classifier is more important than the their parameters tuning, the former should thus deserve more computational budget. In this regard, the bi-level programming formulated in \texttt{BiLO-CPDP} provides better flexibility to achieve such a requirement of fine-grained budget allocation~\cite{ChenLBY18,ZouJYZZL19,LiZZL09,BillingsleyLMMG19,LiZLZL09,Li19,LiCSY19,WuLKZZ19,LiK14,LiFK11,LiKWTM13,CaoKWL12,CaoKWL14,LiKZD15,LiDZZ17,LiKD15,ChenLY18,LiZKLW14,LiFKZ14,LiKWCR12,LiWKC13,CaoKWLLK15,LiCFY19,WuLKZ20,WuKJLZ17,WuLKZZ17,LiDY18,WuKZLWL15,LiDZK15,LiKCLZS12,LiDAY17,LiDZ15,LiXT19,GaoNL19,LiuLC19,LiZ19,KumarBCLB18,Liu0020,CaoWKL11}.
\end{itemize}

\section{Threats to Validity}
\label{sec:threats}

Similar to many empirical studies in software engineering, our work is subject to threats to validity. 

Construct threats can be raised from the experiment uncertainty caused by the learning and optimization. To mitigate this, we have repeated 30 runs for each techniques and compare the techniques using Scott-Knott test~\cite{MittasA13}, supported by Wilcoxon signed-rank test~\cite{Wilcoxon1945IndividualCB} and $A_{12}$ effect size metric~\cite{Vargha2000ACA}. Therefore, whenever we report \textit{``A is better than B"}, we imply that A is indeed statistically better with large effect size. The single metric AUC may also subject to such a threat. However, AUC was chosen mainly due to its parameter-free nature and high reliability as reported in the machine learning community~\cite{LingHZ03}.


Internal threats can be related to the parameter setting, which in our case the key parameter is the time budget for optimization. Indeed, a different budget may affect the result, and therefore we have set a total budget following the state-of-the-practice suggested in the AutoML community~\cite{FeurerKESBH15}, which is reasonable given the required runs. We have also investigated the relative importance of budget allocation between the upper- and lower-level in \texttt{BiLO-CPDP}.

External threats are concerned with whether the findings are generailzable to other projects. To mitigate such, as discussed in~\pref{sec:evaluation}, our 20 projects cover a wide spectrum of the real-world cases with diverse characteristics, each of which was used as the target domain data to be predicted using the other 19 ones as sources.



\section{conclusion}
\label{sec:conclusion}

The choice of combination of transfer learner and classifier along with their hyper-parameter settings have a significant impact to the performance of CPDP model. In this paper, we propose \texttt{BiLO-CPDP}, a tool that is able to automatically develop a high-performance CPDP model for the given CPDP task. Specifically, \texttt{BiLO-CPDP}, for the first time, formulates the automated CPDP model discovery problem from a bi-level programming perspective. In particular, the upper-level optimization routine searches for the right combination of transfer learner and classifier while the lower-level optimization routine optimizes the corresponding hyper-parameters associated with the chosen combination. Furthermore, the hierarchical optimization paradigm allows a more flexible control of the computational budget at both levels. From our empirical study, we have shown that \texttt{BiLO-CPDP}

\begin{itemize}
    \item automatically develops a better CPDP model comparing to 21 state-of-the-art CPDP techniques with hand-crafted combination and reported parameter settings.
    \item overwhelmingly outperforms \texttt{Auto-Sklearn}, a state-of-the-art AutoML tool, and the single-level optimization variant of \texttt{BiLO-CPDP}.
    \item allows software engineers to set more search budget for the upper-level, which significantly boosts the performance.
\end{itemize}

\texttt{BiLO-CPDP} showcases the importance of automatically optimizing the combination of transfer learners and classifiers, together with their parameters. This paves a new way to enable more intelligent parameter optimization and adaptation for CPDP model building. In future, we seek to consider multiple objectives within the bi-level programming and to investigate more precise effects of allocating budget between the two levels. We also plan to further distinguish between the parameters for transfer learner and classifier at the low-level, as it has been shown that the parameter tuning of the former is more important than the latter~\cite{LiXCWT20}.

\section*{Acknowledgement}
K. Li was supported by UKRI Future Leaders Fellowship (Grant No. MR/S017062/1).

\bibliographystyle{IEEEtran}
\bibliography{IEEEabrv,reference}

\end{document}